\documentstyle{article}
\date{March 9, 1994}
\newtheorem{theo}{Theorem}
\newtheorem{defn}[theo]{Definition}

\newtheorem{lemma}[theo]{Lemma}
\newtheorem{prop}[theo]{Proposition}
\newtheorem{coro}[theo]{Corollary}

\def\g{{\scriptstyle {\cal G}}}
\def\nm{\phantom{write something}}
\def\be{\begin{equation}}
\def\ba{\begin{eqnarray}}
\def\ee{\end{equation}}
\def\ea{\end{eqnarray}}
\def\o{\otimes }
\def\bo{\mbox{\,\raisebox{-0.65mm}{$\Box$} \hspace{-4.7mm}
${\scriptstyle\times}$ \/}}
\def\D{\Delta }
\def\1{{\bf 1}}

\def\U{{\cal U}}
\def\k{{\cal k}}

\def\A{{\cal A}}
\def\B{{\cal B}}
\def\G{{\cal G}}
\def\F{{\cal F}}
\def\S{{\cal S}}

\def\R{{\cal R}}
\def\U{{\cal U}}
\def\J{{\cal J}}

\def\ti{\times }
\def\vp{\varphi }

\def\s{\sigma }

\def\a{\alpha }
\def\b{\beta }
\def\c{\gamma }
\def\d{\delta }
\def\e{\epsilon }

\def\k{\kappa }

\def\th{\theta}

\def\vac{|0 \rangle}
\def\t{\tau}

\def\nn{\nonumber}

\def\bt{\bar{\tau}}

\renewcommand{\!}{\hspace*{-2.5mm}}
\newcommand{\ew}{\hspace*{-2mm}}

\newcommand{\CG}[6]{[ \ew \begin{array}{lll} {\scriptstyle #1} \!
  & {\scriptstyle #2} \! & {\scriptstyle #3} \ew \\[-2mm]
{\scriptstyle
  #4} \! & {\scriptstyle #5}\! & {\scriptstyle #6} \ew\end{array}  ]}
\newcommand{\SJS}[6]{ \{ \ew \begin{array}{lll} {\scriptstyle #1} \!
&
  {\scriptstyle #2} \! & {\scriptstyle #3} \! \\[-2mm]{\scriptstyle
#4}
  \! & {\scriptstyle #5} \! & {\scriptstyle #6} \ew \end{array}  \}
_q }
\newcommand{\pr}[1]{\ _{#1}}

\begin{document}
\begin{titlepage}
\title{Combinatorial Quantization of the Hamiltonian Chern-Simons
Theory I}
\author{{\sc Anton Yu. Alekseev}
\thanks{On leave of absence from Steklov Mathematical Institute,
Fontanka 27, St.Petersburg, Russia; e-mail:
alexeiev@rhea.teorfys.uu.se}
\thanks{Supported by Swedish Natural Science Research Council (NFR)
under the contract F-FU 06821-304 and by the Federal Ministry of
Science and Research, Austria}
\\Institute of Theoretical Physics, Uppsala University,
\\ Box 803 S-75108, Uppsala, Sweden \\[4mm]
{\sc Harald Grosse}
\thanks{Part of project P8916-PHY of the 'Fonds zur F\"{o}rderung
der wissenschaftlichen Forschung in \"{O}sterreich'; e-mail:
grosse@pap.univie.ac.at}
\\ Institut f\"ur  Theoretische Physik,
Universit\"at Wien, Austria\\[4mm]
{\sc Volker Schomerus \thanks{Supported in
part by DOE Grant No DE-FG02-88ER25065; e-mail:
vschomer@husc. harvard.edu}} \\
Harvard University, Department of Physics \\ Cambridge, MA 02138,
U.S.A.}
\maketitle \thispagestyle{empty}

\begin{abstract}
Motivated by a recent paper of Fock and Rosly \cite{FoRo} we describe
a mathematically precise quantization of the Hamiltonian Chern-Simons
theory. We introduce  the Chern-Simons theory on the lattice
which is expected to reproduce the results of the continuous theory
exactly.
The lattice model enjoys the symmetry with respect to a quantum
gauge group. Using this fact we construct the algebra of observables
of the Hamiltonian Chern-Simons theory equipped with a *- operation
and a positive inner product.
\end{abstract}
\vspace*{-15.5cm}
\hspace*{9cm}
{\large \tt HUTMP 94-B336} \\
\hspace*{9cm}
{\large \tt ESI 79 (1994)}         \\
\hspace*{9cm}
{\large \tt UUITP 5/94}   \\
\hspace*{9cm}
{\large \tt UWThPh-1994-8}   \\
\hspace*{9cm}
{\large \tt hep-th 9403066}
\end{titlepage}

\section{Introduction}
\def\tt{\tilde{\tau}}
\setcounter{equation}{0}

Quantization of  the Chern--Simons theory in 3 dimensions
has been attracting attention and efforts of  many physicists
and mathematicians during the last 5 years. The most spectacular
results obtained in this way are the construction of knot invariants
\cite{Wit1} and exact solution of $2+1$ dimensional gravity
\cite{Wit2}. Being a 3 dimensional topological field theory,
the Chern-Simons model is intimately related to the
Wess-Zumino-Novikov-Witten (WZNW) model of conformal field theory
in 2 dimensions and to  quantum groups (one may regard
a quantum group as a 1 dimensional quantum system). Actually,
it is this hierarchy of systems in different dimensions which
makes the Chern-Simons theory solvable. The relation between
CS and WZNW models allowed to  evaluate partition functions
and correlators \cite{Wit1} using the methods of conformal field
theory. The importance of the CS theory is a motivation to look for
different approaches. Among the others the perturbation
theory \cite{AS} and exact evaluation of  the functional
integral by means of localization formulae \cite{Ger} should be
mentioned.

In this paper we develop an approach to the CS model based on its
relation to the theory of quantum groups.  One of the advantages of
the quantum
group approach is that we deal with finite dimensional objects only.
As a consequence,  one can  represent the answers in terms of
finite sums , whereas in other approaches the final result usually
has an integral form. It makes the quantum group approach helpful in
dealing with topology of 3D manifolds
\cite{TuVi},\cite{DJN},\cite{KaSc}
and in knot theory \cite{Piun}, \cite{Cart}.
The main idea is to simulate the Chern-Simons theory on
the lattice in such a way that  partition functions and correlators
of the lattice model coincide with those of the continuous CS model.
It is important that the lattice model enjoys the
gauge symmetry with respect to the quantum group.
It is worth mentioning that  a lattice simulation of the CS
model  has been
suggested in \cite{Bou}.
The drawback of this model is the absence of  gauge symmetry.
The gauge symmetry may be restored if one uses the proper
combinatorial
description of the moduli space of flat connections \cite{FoRo}.

Let us briefly characterize the content of each Section. In Section
2
we review the main facts concerning the
CS theory in Hamiltonian approach and introduce the
combinatorial (or lattice) description following \cite{FoRo}.
Section 3 is devoted to the quantum gauge group in
the lattice model. In Section 4 we generalize the concept
of  lattice gauge fields to the case of a
quantum gauge group. The algebra of observables corresponding
to the Hamiltonian CS theory appears in Section 5 equipped with a
*-operation. We describe the Hermitian inner product in the algebra
of observables and prove the positivity theorem in Section  6.
In section 7 we generalize the theory to
weak quasi-Hopf algebras \cite{MSIII}. In this way we can
deal with universal enveloping algebras at  roots of
unity by using a procedure called ``truncation''. The basic
definition of weak quasi-Hopf algebras and the truncation are
reviewed in section 7.1.
An outlook at the end of the paper is devoted
to the possible perspectives of
the lattice approach.

The basic technique which we use in this paper is the theory of
quantum groups. Section 2 is supposed to play
the role of the physicist oriented introduction.
In the remaining part  we assume that the reader is familiar with
standard definitions and properties of such objects as
co-multiplication and
$R$-matrix.

We want to stress that the construction we propose in the following
assigns a Chern-Simons type model to every pair of a marked
Riemann surface and a weak quasi-Hopf algebra. The latter is not
necessarily given by
a truncated quantized enveloping algebra of some simple Lie
algebra. It has been shown recently \cite{Sch3} that a weak
quasi-Hopf
symmetry can be constructed for every low dimensional quantum
field theory. Combining this result with the considerations
below, one assigns a Chern Simons type model to every low dimensional
quantum field theory.
It remains unclear whether these generalized Chern Simons
theories are related to certain moduli spaces
in the same way as standard Chern Simons theories are related to
the moduli space of flat connections.

\section{Physical motivations}
\setcounter{equation}{0}

In this paper we study the problem of quantization of
the Chern-Simons theory within the Hamiltonian approach.
The moduli space of flat connections on a  Riemann surface
appears as a phase space for this model \cite{Wit1}. Let us briefly
remind the definition and general features of the CS system.

\subsection{Chern - Simons model}
The Chern-Simons  theory is a gauge theory in 3 dimensions
(in principle the CS term exists in any  odd dimension). It is
defined by the action principle
\begin{equation}
CS(A)=\frac{k}{4\pi}Tr\int_{M} ( AdA  + \frac{2}{3} A^{3} )\ \ .
\label{CSaction}
\end{equation}
Here $M$ is a 3-dimensional (3D) manifold, $k$ is a positive integer
and  the gauge field $A$ takes
values in some semisimple Lie
algebra ${\g}$
\begin{equation}
A = A^{a}_{i}t^{a}dx_{i}\ \ .                  \label{3Dcon}
\end{equation}
The generators $t^{a}$ form a basis in ${\g}$ and satisfy
the commutation relations
\begin{equation}
[ t^{a} , t^{b} ] = f^{ab}_{c} t^{c}\ \ .       \label{Liecomm}
\end{equation}

In this paper we  concentrate on the very particular
version of the CS theory when it has a Hamiltonian
interpretation. Suppose that the manifold M locally
looks like a cylinder $\Sigma\times R$ (Cartesian product of  a
Riemann  surface $\Sigma$ and a segment of the real line). Then
we may
choose the direction parallel to the real
line $R$ to be the time direction. Two space--like components
of the gauge field $A$ become dynamical variables and
we shall often denote by $A$ the two component gauge field
on the  surface $\Sigma$. As usual, the time-component
$A_{0}$  becomes a Lagrangian multiplier.
After the change of variables the action (\ref{CSaction})
acquires the form
\begin{equation}
S =\frac{k}{4\pi}Tr\int (- A \partial_{0} A + 2A_{0} F) dt\ \ ,
\label{HamCS}
\end{equation}
where the first term is just like $\int pdq$ and the
second term introduces the
 first class constraint
\begin{equation}
F = dA + A^{2} = 0\ \ .                          \label{F=0}
\end{equation}
The first term in (\ref{HamCS}) determines
the Poisson brackets (PB) of dynamic variables. In particular, the
Poisson
bracket of the constraints (\ref{F=0}) may be easily calculated:
\begin{equation}
\{F^{a}(z_{1}),F^{b}(z_{2})\}=\frac{2\pi}{k}f^{ab}_{c} F^{c}(z_{1})
\delta^{(2)}(z_{1} - z_{2})\ \ .                \label{PBconstr}
\end{equation}
As one expects, the constraints (\ref{F=0}) generate gauge
transformations
\begin{equation}
A^{g} = g^{-1}Ag+g^{-1}dg\ \ .             \label{Ggtr}
\end{equation}
Thus, the phase space of the Hamiltonian CS theory is a
quotient of the space $\Im$ of flat connections (\ref{F=0})
over the gauge group $\Sigma G$ (\ref{Ggtr}). We see that the moduli
space (we shall often refer to the moduli space of flat
connections as to the moduli space) appears to be a phase space  of
the CS theory on the cylinder. The action principle (\ref{HamCS})
provides canonical Poisson brackets on the moduli space. An
efficient description of
this PB was given in \cite{FoRo} (see also subsection 2.3).

\subsection{Wilson lines and marked points}

We continue our brief survey of the CS theory and
consider possible observables. The CS model enjoys
two important symmetries: gauge symmetry and the symmetry
with respect to diffeomorphisms. The reparametrization
symmetry appears due to the geometric nature of the action
(\ref{CSaction}) which is written in terms of
differential forms and   automatically invariant
with respect to diffeomorphisms of  the  manifold $M$. It is natural
to require that the observables in the CS model respect the
invariance properties of the theory.  Some observables of this
type  may be constructed starting from the following
data. Let us choose the closed contour $\Gamma$ in $M$ and
a representation $I$ of the algebra $\cal G$. Apparently the
following functional of the gauge field A
\begin{equation}
W_{I}(\Gamma) = Tr_{I} P exp (\int_{\Gamma} A^{I})      \label{Wl}
\end{equation}
is invariant with respect to both gauge and reparametrization
symmetries. Usually the contour $\Gamma$ is called a Wilson line
and the
expression (\ref{Wl})  is called
a Wilson line observable. The connection $A^{I}$ is  equal to
\begin{equation}
A^{I} = A^{a} T_{a}^{I}\ \ ,             \label{AI}
\end{equation}
where matrices $T_{a}^{I}$
represent the algebra $\g$ in the representation $I$.

In the Hamiltonian formulation we may choose two special classes
of Wilson lines: vertical and horizontal.

We call a Wilson line horizontal if it lies on an equal
time surface. The observable corresponding to a
horizontal Wilson line is a functional of the two-dimensional
gauge field and
after quantization it becomes a physical operator.

The Wilson line is called vertical if the contour $\Gamma$ is
parallel to the time axis. In the Hamiltonian picture we
do not actually control the fact that vertical Wilson lines are
closed. They come
from the past, go through the zero-time surface
and disappear in the future.
The vertical Wilson line is characterized by the representation
$I$ and the point $z$ where it intersects the Riemann surface
$\Sigma$. The choice of the time axis produces a big difference in
the role of horizontal and vertical Wilson lines in the theory.
Vertical Wilson lines {\em do not} correspond to observables
in the Hamiltonian formulation. Instead, they change the Hamiltonian
system (\ref{HamCS}) so that both $\int pdq$ and the constraint
get modified.

As a preparation for analyzing vertical Wilson lines let
us remind the correspondence between co-adjoint orbits and
representations of semi-simple Lie groups. Let $I$ be a
representation of the Lie algebra ${\g }$:
\begin{equation}
[ T^{I}_{a} , T^{I}_{b} ] = f^{ab}_{c} T^{I}_{c}\ \ .     \label{Grep}
\end{equation}
One may ask the question what is the underlying classical
system corresponding to the quantum algebra (\ref{Grep}).
The first idea is to simulate the commutation relations by
the Poisson brackets:
\begin{equation}
\{T_{a},T_{b}\}=f^{ab}_{c}T_{c}\ \ ,    \label{KPB}
\end{equation}
where $T_{a}$   are commuting coordinates (on the dual space to
the Lie algebra). It seems that the information about the
representation disappeared in the formula (\ref{KPB}).
However, this is not quite true. The PB
(\ref{KPB}) is always degenerate. At this point it is useful to
introduce a coordinate matrix
\begin{equation}
T =T_{a}t^{a}\ \ ,
\end{equation}
where $t^{a}$ form a basis in the fundamental representation.
In order to make the bracket (\ref{KPB}) nondegenerate one
should fix eigenvalues of the matrix $T$:
\begin{equation}
T = g^{-1}Pg\ \ .                      \label{Orb}
\end{equation}
A conjugation class (\ref{Orb}) is also called a co-adjoint orbit.
Now we have a diagonal matrix $P$ which parametrizes the set of
orbits. Quantization of a particular orbit
leads to an irreducible representation of the Lie
algebra. The highest weight  $w_{I}$ of the representation $I$
represented as a diagonal matrix in the fundamental representation is
related to  $P$ by
\begin{equation}
w_{I} = P - \rho\ \ ,                          \label{shift}
\end{equation}
where $\rho$ is a half sum of positive roots of $\g$.

The information about the bracket (\ref{KPB}) on the orbit
may be encoded in the action
\begin{equation}
S_{I}(g)=Tr\int Pdgg^{-1}\ \ .                     \label{Gaction}
\end{equation}
The action  (\ref{Gaction}) is called a  geometric action because
it originates from the method of geometric quantization.
The family of systems (\ref{Gaction}) is parametrized by the
set of representations of ${\g}$ which may be obtained upon
quantization.

The essence of the quantization procedure  for the action
(\ref{Gaction})
is the following formula for a Wilson line observable:
\begin{equation}
W_{I}=\int e^{i(S_{I}(g) + \int A_{0}^{a}T^{a} dt)} Dg\ \ .
\label{WLn}
\end{equation}
The ordered exponent and trace are provided by the functional
integral automatically.

Using the formula (\ref{WLn}), one may treat the CS correlator
with $n$ vertical Wilson lines inserted
\begin{equation}
Z_{k}(I_{1},\dots,I_{n}) = \int DA e^{iCS(A)}
W_{I_{1}}\dots
W_{I_{n}}                        \label{Wilpart}
\end{equation}
as an expression where the gauge field is still classical, whereas
some modes corresponding to the matrices $T_{i}$ are already
quantized.
The original functional integral would be
\begin{equation}
Z = \int DA Dg_{1}\dots Dg_{n} e^{iS^{tot}}\ \ .  \label{totpart}
\end{equation}
The action $S^{tot}$ is defined by the formula
\begin{equation}
S^{tot} =CS(A)+\sum_{i=1}^{n}(S_{I_{i}}(g_{i}) +
Tr\int A_{0}(z_{i})T_{i}dt)\ \ .                 \label{totaction}
\end{equation}
Here the first term coincides with the standard Chern-Simons action,
the second term consists of two parts. The first part collects
auxiliary geometric actions for each Wilson line, the second part
represents contributions of the Wilson lines into the  CS partition
function (\ref{Wilpart}).

We have reformulated the Hamiltonian Chern-Simons model with
vertical Wilson lines as a theory of the 2D gauge field $A$ interacting
with a set of finite dimensional systems with coordinates
$T_{i}$ localized at the points $z_{i}$. As in the case of the pure
CS theory,
the Hamiltonian (\ref{totaction}) is equal to zero.  The action of
the modified
system may be rewritten as
\begin{equation}
S^{tot} = Tr (-\frac{k}{4\pi}\int A \partial_{0} A +
 \sum_{i=1}^{n} P_{i}dg_{i}g_{i}^{-1} ) +
 Tr \int A_{0} (\frac{k}{2\pi}F +
\sum_{i=1}^{n}T_{i}\delta(z-z_{i}))\ .
                                 \label{Hamtot}
\end{equation}
The first term in (\ref{Hamtot}) is of the type $\int pdq$ of the
Hamiltonian system. It is responsible for the Poisson brackets of
dynamical variables. The second term gives the modified constraint
\begin{equation}
\Phi(z)=F+\frac{2\pi}{k} \sum_{i=1}^{n}T_{i}\delta(z-z_{i})=
0\ \ .
                                      \label{Modconstr}
\end{equation}
The constraints (\ref{Modconstr}) satisfy the same algebra
(\ref{PBconstr}) as in the pure CS theory. They generate
gauge transformations for the gauge field $A$ and conjugations
for the variables $T_{i}$:
\begin{equation}
A^{g} = g^{-1}Ag+g^{-1}dg \ \ , \ \  T_{i}^{g} = g(z_{i})^{-1}T_{i}
g(z_{i}).
                                           \label{ModGtr}
\end{equation}

Now the phase space of the Hamiltonian Chern-Simons theory
with vertical Wilson lines may be described. First we mark $n$
points $\{z_{i}\}$ on the Riemann surface $\Sigma$ of genus $g$.
Each point  is equipped with a representation $I_{i}$ of the algebra
$\g$
(and corresponding orbit $O_{i}$). One can choose a  subspace
$\Im(I_{1},\dots ,I_{n})$ defined by the constraint
(\ref{Modconstr})
in the Cartesian product $C_{g}\times O(I_{1})\times \dots  \times
O(I_{N})$
of the space of all connections $C_{g}$  on the Riemann surface and
co-adjoint orbits attached to the marked points.
The subspace $\Im(I_{1},\dots ,I_{n})$  is a natural analogue
of the space of flat connections. It is invariant with respect
to gauge transformations (\ref{ModGtr}). The quotient of
$\Im(I_{1},\dots ,I_{n})$ by the action of the
gauge group may be called the moduli space ${\cal M}_{g,n}(\g)$ of
flat
connections on the Riemann surface of genus $g$ with $n$ marked
points.
The moduli space ${\cal M}_{g,n}(\g)$ inherits the Poisson structure
from the CS theory. This structure is the subject of the next
subsection.

Let us finish by a short remark concerning the structure
of ${\cal M}_{g,n}(\g)$ and the theory of orbits. Choosing a small
loop
$\Gamma_{i}$ surrounding the marked point $z_{i}$, one can define the
monodromy matrix (or parallel transport)  $M_{i}$ along this path.
It is easy to check that if $A$ and $\{T_{i}\}$ satisfy
(\ref{Modconstr}),
the monodromy matrix $M_{i}$ may be diagonalized
by conjugation of the exponent of $P_{i}$
\begin{equation}
M_{i} = h^{-1}exp(\frac{2\pi}{k} P_{i})h\ \ .  \label{CExporb}
\end{equation}
In the quantum case there is a one-loop correction in this formula
which adds the dual Coxeter number $h^{*}$  to the parameter $k$.
The correct quantum version of (\ref{CExporb}) reads as
\begin{equation}
M_{i} = h^{-1}exp(\frac{2\pi}{k+h^{*}} P_{i})h\ \ . \label{QExporb}
\end{equation}
Thus, the monodromy matrix  belongs
to the exponentiated orbit assigned to the corresponding marked
point. Formulae (\ref{CExporb}) and (\ref{QExporb}) characterize
elementary monodromies. They will be quite helpful for quantization
of ${\cal M}_{g,n}(\g)$.

\subsection{Combinatorial description and Exchange relations}

In the previous subsection we have found  that the moduli space
${\cal M}_{g,n}(\g)$
appears as a phase space in the Chern-Simons theory with vertical
Wilson lines. In order to quantize the theory we need the Poisson
structure on
${\cal M}_{g,n}(\g)$. In principle one may proceed starting from the
2D gauge fields  with the Poisson bracket

\begin{equation}
\{A_{i}^{a}(z_{1}),A_{j}^{b}(z_{2})\}=-\frac{2\pi}{k} \delta^{ab}
\epsilon_{ij} \delta^{(2)}(z_{1}-z_{2})\ \ .   \label{APB}
\end{equation}
The quantization of the relations (\ref{APB}) is straightforward
\begin{equation}
[A_{i}^{a}(z_{1}), A_{j}^{b}(z_{2})]=-\frac{2\pi}{k} \delta^{ab}
\epsilon_{ij} \delta^{(2)}(z_{1}-z_{2})\ \ .   \label{ACom}
\end{equation}
Observables may be described as gauge invariant
functionals of $A$, where the constraint (\ref{Modconstr}) is
imposed.
In this approach we deal with the representations of the
infinite dimensional algebra (\ref{ACom}) and construct the
quantum mechanics corresponding to the CS system starting from
the field theory of the gauge field $A$. It is  a motivation
to look for another approach. A recent  progress
\cite{FoRo} in this direction allows to reformulate the problem.
The idea is to simulate the Riemann surface
by the oriented fat graph drawn on it. Dealing with a fat
graph one uses ribbons
instead of strings. It means that the cyclic order
of links incident to a given vertex is fixed.

Suppose that we have a gauge field $A$ on the Riemann surface
and  a graph drawn on it. We assume that the surface is
divided by the graph into plaquettes so that any plaquette is
contractible. The graph should be chosen in such a way  that the
number of marked points inside each plaquette does not exceed one.
The gauge field defines a parallel transport along each
link of the graph. Let us enumerate the vertices of the graph by
letters $x,y,z\ldots$ and  the links by $i,j,k\dots$. It is
convenient
to introduce notations $s(i)$ and $t(i)$ for the end-points of the
link $i$.  The parallel transport corresponding to this link may be
written as an ordered exponent
\begin{equation}
U(i) = P exp (\int_{s(i)}^{t(i)}A)\ \ .    \label{GRCon}
\end{equation}
As for the Wilson lines, one may introduce the set of matrices
$U^{I}(i)$
\begin{equation}       \label{UI}
U^{I}(i) = P exp (\int_{s(i)}^{t(i)}A^{I}).
\end{equation}
using the gauge field in different representations (\ref{AI}).

Some information about the connection $A$ is encoded in the
link variables (\ref{GRCon}). The question is whether this
information  is sufficient to reconstruct the
moduli space ${\cal M}_{g,n}(\g)$?  The answer is obviously positive.
To recover the moduli space we should factorize over residual gauge
transformations and take into account  the flatness condition.
Gauge transformations act on the graph connections (\ref{GRCon})
as follows
\begin{equation}
U(i)^{h} = h(s(i))^{-1} U(i) h(t(i)).        \label{Latgauge}
\end{equation}
It is remarkable that the gauge group becomes effectively
finite dimensional because only values of $h$ in graph vertices
enter into (\ref{Latgauge}).

The condition of flatness may be simulated using the properties of
the monodromy from the previous Subsection. We form the monodromy
for each plaquette and constrain it by the condition
(\ref{CExporb}).
If there is no marked point inside the plaquette, the monodromy is
simply equal to the identity.

It is proved \cite{FoRo} that factorizing the
space of flat graph connections over graph gauge transformations
one obtains the same space ${\cal M}_{g,n}(\g)$! Moreover, the
Poisson
structure on the space of graph connections leading to the standard
Poisson structure on ${\cal M}_{g,n}(\g)$ is known. In this approach
the moduli space is represented as a quotient of the finite
dimensional space over the finite dimensional group.
This is the reason to call it
combinatorial description of the moduli space.

One may think that the Poisson brackets for graph connections
are defined uniquely since $U(i)$ are functionals of the gauge
field $A$ and the Poisson brackets for $A$ are fixed by (\ref{APB}).
In fact, this is not correct. The reason is the $\delta$- function
singularity in the brackets (\ref{APB}). Calculating the PB of
two matrices $U(i)$ and $U(j)$, where the links $i$ and $j$ have
at least one common end-point, one has to resolve the  singularity
appearing at this very point. There is no canonical way of
resolution. In general the Jacobi identity breaks
down after  regularization. It means that one can't construct
the Poisson bracket applicable for both local gauge
fields $A(x)$ and arbitrary link variables $U(i)$.
However, it is possible to introduce  meaningful
Poisson brackets  for link variables $U(i)$.
So, the brackets for $A(x)$ and for
$U(i)$ are to some extent independent from each other.
One calls them consistent if they give the same answer
for Poisson brackets of gauge invariant variables.
In other words, $A$- and $U$-bracket  may differ
in nonphysical sector but they coincide when we
restrict ourselves to physical gauge invariant
observables.

In order to simplify the analysis of the $U$-bracket,
we  remind here  some standard definitions and properties
of quadratic Poisson brackets and Exchange algebras.

Suppose that we have a Poisson bracket defined on the matrix group.
The simplest form of such bracket is quadratic in matrix elements
\begin{equation}
\{ U^{1} , U^{2} \} = U^{1}U^{2}r \ \ ,      \label{Qbr}
\end{equation}
where we have used tensor notations
\begin{equation}
U^{1} = U \otimes id, U^{2} = id \otimes U   \label{Tennot}
\end{equation}
and the matrix $r$ is defined in the tensor product of two
vector spaces. To ensure the Jacobi identity, the following
condition on the matrix $r$ must be satisfied
\begin{equation}
[ r_{12}, r_{23} ] + [ r_{12} , r_{13} ] +
[ r_{13} , r_{23} ] = 0\ \  .
\label{CYB}
\end{equation}

The constraint (\ref{CYB}) is the classical Yang-Baxter equation.
Fortunately, for any simple Lie algebra we know two solutions to
the very complicated equation (\ref{CYB}). Usually they are called
$r$ and $r'$ and look as follows
\begin{equation}
r = \sum_{i} h^{i}\otimes h^{i} + 2\sum_{\alpha}t^{\alpha}
\otimes t^{-\alpha} \ \ ,                 \label{r+}
\end{equation}
\begin{equation}
r'= \sum_{i} h^{i}\otimes h^{i}+2\sum_{\alpha}t^{-\alpha}
\otimes t^{\alpha} \ \ .                 \label{r-}
\end{equation}
Here the sum in the first term runs over the set of simple roots
and in the second  over the set of positive roots.

The  Poisson algebra (\ref{Qbr}) may be quantized if  one
knows a one parameter
family of solutions $R(h)$ of the quantum Yang-Baxter equation
\begin{equation}
 R_{12} R_{13} R_{23}   =   R_{23} R_{13} R_{12}    \label{QYB}
\end{equation}
with given asymptotics
\begin{equation}
R = 1 + hr + \ldots   \ \ .     \label{Rr}
\end{equation}
Using $R(h)$ we quantize the bracket (\ref{Qbr}) in the
following way
\begin{equation}
 U^{1}  U^{2}  = U^{2} U^{1} R\ \ .      \label{QEx}
\end{equation}
We shall denote solutions of the quantum Yang--Baxter
equation corresponding to the classical $r$-matrices
(\ref{r+},\ref{r-})
by $R$ and $R'$. It is assumed everywhere in the text that
both of them depend on the deformation parameter $h$.  It is worth
mentioning that
\begin{equation}
R' = P R P\ \ ,           \label{PRP}
\end{equation}
where $P$ is a permutation matrix in the tensor product of two
copies of the same vector space. It is important that  along with
$R$- matrices in the fundamental representation there exists
a family of $R$- matrices parametrized by pairs $(I,J)$ of finite
dimensional representations of ${\g}$ so that $R^{IJ}$ act
in the tensor product of the representation
spaces of $I$ and $J$ and  equation (\ref{QYB})
holds for any triple of representations $(I,J,K)$
corresponding to the indices 1,2,3.

In the class of regularizations suggested in \cite{FoRo} the Poisson
brackets  of graph connections are
quadratic. Let us first describe these brackets and then
quantize.

To fix Poisson brackets for graph connections one must choose
some particular regularization of the singularity in (\ref{APB}).
In practice it means that each vertex  should be equipped  with a
classical $r$-matrix (in the quantum case -- quantum
$R$- matrix)  from a certain equivalence class.
Roughly speaking the class is defined by the choice of the
deformation parameter $h$. One also should fix a linear order
of incident links at each vertex in addition to the natural cyclic
order.
The latter may be done by putting  a little eyelash at each
graph vertex. The eyelash determines from where we enumerate links
coming to the vertex. Concerning the choice of $r$-matrices, we shall
restrict ourselves to the particular regularization so that  the
$r$-matrices assigned to all vertices of the graph
coincide.

There are three rules which determine the
structure of the Poisson algebra of graph connections:

1) The Poisson bracket of any matrix elements of two
parallel transport matrices
corresponding to links which have no common end-points vanishes:
\begin{equation}
\{ U(i)^{1}, U(j)^{2}\}=0 \ \ .       \label{PEx1}
\end{equation}
This condition brings locality into the definition of the
Poisson bracket. Indeed, the original bracket for 2-dimensional
connections had a support at coincident points. If the links
have no common
end-points, they do not intersect at all. It means that the bracket
of the corresponding  matrix elements should vanish if we want to
reproduce the continuous theory.

2) For the matrix elements of the same matrix we have
\begin{equation}
\{ U(i)^{1}, U(i)^{2}\}=\frac{2\pi}{k} (rU(i)^{1}U(i)^{2}
-U(i)^{1}U(i)^{2}r')\ \ .      \label{PEx2}
\end{equation}

3) If the links have one common end-point, the Poisson bracket
acquires the form
\begin{equation}
\{ U(i)^{1}, U(j)^{2}\}=\frac{2\pi}{k} U(i)^{1} U(j)^{2}r\ \  ,
      \label{PEx3+}
\end{equation}
if the link $i$ is elder than the link $j$ in the  clock-wise
order starting from the eyelash (we can express it as $i<j$) and
\begin{equation}
\{U(i)^{1}, U(j)^{2}\}=-\frac{2\pi}{k} U(i)^{1} U(j)^{2}r'
\label{PEx3-}
\end{equation}
otherwise.

{}From the definitions of the Poisson algebra on the space of graph
connections we learn that the deformation parameter in this theory
is equal to $h=\frac{2\pi}{k}$. Actually, this formula is correct
only semiclassicaly. In the Hamiltonian formulation the
Chern-Simons integral is Gaussian. So, we should worry only about
possible one-loop corrections. Depending on the definition of the
integration measure in the Chern-Simons functional integral, one gets
different results for the renormalized value of the parameter $k$.
In the standard scheme which we follow in this paper,
$k$  receives a one loop correction equal to the dual Coxeter
number $k\rightarrow k+h^{*}$. So, the correct value of the
deformation
parameter is $h=\frac{2\pi}{k+h^{*}}$.

Another one loop effect which shows up in the Chern-Simons theory
is the framing anomaly (see e.g. \cite{WBN}). It appears that
in order to define the renormalized CS model one should fix
a frame in the 3D manifold $M$ and replace Wilson lines by
ribbons. Physical correlation functions depend on the
choice of the framing. The exact form of this dependence is
governed by the framing anomaly. In the Hamiltonian version
of the CS theory we always have a prefered framing invariant
with respect to shifts along the time axis. We can
stick to this framing from the very beginning and in this
way we do not trace the framing anomaly in the Hamiltonian
approach.

At this point we want to stress
that the information about the one loop correction of $k$ is the only
external data which we bring into the scheme of combinatorial
quantization.
Further steps are quite independent of the Lagrangian formulation
of the theory and give a selfconsistent approach to the Chern-Simons
model.

The relations (\ref{PEx3+}) and (\ref{PEx3-}) are written
for the situation when both oriented links point towards their
common
end-point. All the other relations may be derived following the rules,
if we assume that for the same link taken with two different
orientations the corresponding link variables are inverse to each
other:
\begin{equation}
U(i) U(-i) = U(-i) U(i) = 1\ \ .            \label{+-}
\end{equation}
For example, if  $i<-j$ and the link $j$ starts from the common
end-point the Poisson bracket
(\ref{PEx3-}) will be modified
\begin{equation}
\{ U(i)^1, U(j)^{2}\} =-U(j)^{2}rU(i)^{1}\ \ .      \label{PEx3+'}
\end{equation}
So we have described the Poisson algebra for
the gauge field on the graph. Now the problem
of quantization is in order. As we discussed,
quadratic $r$-matrix Poisson brackets admit
straightforward quantization. Let us list
the corresponding quantum formulae:
\begin{equation}
U(i)^{1}U(j)^{2}=U(j)^{2}U(i)^{1}\ \ , \label{Ex1}
\end{equation}
for links $i$ and $j$ which have no intersection points;
\begin{equation}
U(i)^{1}U(i)^{2}=RU(i)^{2}U(i)^{1}(R')^{-1}\ \ , \label{Ex2}
\end{equation}
for the matrix elements of the same matrix;
\begin{equation}
U(i)^{1}U(j)^{2}=U(j)^{2}U(i)^{1}R,  \label{Ex3+}
\end{equation}
for two links which have a common target when $i<j$ and
\begin{equation}
U(i)^{1}U(j)^{2}=U(j)^{2}U(i)^{1}(R')^{-1}  \label{Ex3-}
\end{equation}
for $i>j$.
The quantum algebra defined by the relations (\ref{Ex1}-
\ref{Ex3-}) may be treated as a noncommutative analogue
of a lattice gauge field. As we see, the lattice emerges
naturally in this approach. Moreover, we do not know how
to get rid of it  in this type of noncommutative gauge models.
It may happen that the lattice formulation is dictated by
the noncommutative nature of the gauge field algebra.

The next question concerns
the generalization of the gauge symmetry
to the noncommutative gauge theory. As in any lattice gauge theory,
gauge transformations act at the lattice vertices:
\begin{equation}       \label{UhUh}
U(i)^{h}=h(s(i))^{-1}U(i)h(t(i))\ \ .
\end{equation}
It is easy to check that these transformations do not preserve
the exchange relations for $U$'s unless we assume that
$h$ entering (\ref{UhUh}) are also noncommutative. More exactly,
$h(x)$ and $h(y)$ commute for $x$ and $y$ being different
vertices
\begin{equation}
h(x)^{1}h(y)^{2}=h(y)^{2}h(x)^{1}
\end{equation}
but form a (not co-commutative) Hopf algebra at each vertex.

The fact that the noncommutative gauge algebra
is invariant with respect to the quantum group
valued gauge transformations may be expressed
also in the following way. We can treat matrix
indices of $U(i)$ as indices of the fundamental
representation of the corresponding quantized
universal enveloping algebra. If we consider
the Chern-Simons theory of $\g$-valued gauge
fields with coefficient $k$ in the action,
the quantum symmetry is an algebra $U_{q}(\g)$
for $q$ being equal to $exp(2\pi i/(k+h^{*}))$.

Here one can change the point of view and
try to construct the noncommutative gauge
fields  starting from some symmetry algebra
placed at the lattice sites. It may be
a quantum group but  one can choose also some
other symmetry algebra. In particular, choosing
the nondeformed Lie algebra $\g $ one should
recover the standard two dimensional
lattice gauge theory. In this paper we
explore the approach based on the symmetry
algebra and find that the gauge theory may
be reconstructed if the symmetry algebra is
endowed with co-multiplication. The latter
means that one can construct tensor products
of representations of the symmetry algebra
and decompose them into irreducible ones.
In more mathematical language it means that
the symmetry algebra is considered as a Hopf
algebra (see Section 3) or as a quasi-Hopf
algebra (see Section 7).

Down to earth, along with the
matrix $U(i)$ in the fundamental representation of the
symmetry algebra we introduce a bunch of matrices
for any representation as in formula (\ref{UI}).
It is not difficult to generalize
quadratic
exchange relations for this case. For example, instead of
(\ref{Ex3+})
we get
\begin{equation}
U^{I}(i)^{1}  U^{J}(j)^{2}  = U^{J}(j)^{2} U^{I}(i)^{1}R^{IJ} .
\label{ExIJ}
\end{equation}
Matrices $U^{I}(i)$ are not independent.
They form a closed algebra so that
a product  of any two matrix elements may be decomposed into a linear
combination of matrix elements. Formula (\ref{+-}) gives a simplest
example of  such relations. The algebra  of
matrix elements of $U^{I}(i)$ is closely related
to the algebra of functions on the finite dimensional group $G$. The
structure constants $[\dots ]$ which appear in the decomposition
\begin{equation}
g^{I}_{ab} g^{J}_{cd} =  \sum  \overline{ \CG{I}{J}{K}{a}{c}{e}^{\a}}
\  g^{K}_{ef} \ \CG{I}{J}{K}{b}{d}{f}^\a
 \label{CG}
\end{equation}
for the  matrix elements of $G$ are usually called Clebsch-Gordan
coefficients.
They are defined as invariant tensors in the triple tensor product of
representations
$I$,$J$ and $K$ and parametrized by an integer $\alpha$.
In  formula (\ref{CG})
the summation over  $e,f,K$ and $\alpha$ is assumed.
Generalizing such relations, the Hopf (or quasi-Hopf)
structure of the symmetry algebra enters in practice.
The algebra of
$U^{I}(i)$ looks exactly like (\ref{CG}) but
the structure constants $C$ must be replaced by the
Clebsch-Gordan coefficients for the corresponding quantum algebra.

Let us give a simple example to clarify the definition of  the
exchange algebra for graph connections. We pick up an
elementary plaquette on the Riemann surface and enumerate
the  links from 1 to $s$ in the counter clock-wise order. It is
convenient to choose the
orientations so that all the arrows are also directed  against the
clock rotation. We choose the eyelashes at all vertices to
be directed inside the plaquette. Under these conditions
link variables $U(1)\dots U(s)$ may be treated  independently of the
rest of the graph. The corresponding exchange
algebra looks as follows:
\begin{equation}
U(i)^{1}  U(i)^{2}=RU(i)^{2}U(i)^{1}(R')^{-1}  \label{LC1}
\end{equation}
for any link $i$;
\begin{equation}
 U(i)^{1}  U(i+1)^{2}  = U(i+1)^{2} R^{-1} U(i)^{1}, \label{LC2}
\end{equation}
where we assume that by definition $U(s+1)=U(1)$. As usual, the
matrix elements of $U(i)$ and $U(j)$ commute if
$i$ and $j$ have no common end-points.

Actually, the graph connection algebra does not know if there is  a
piece of surface
inside the plaquette or, perhaps, there is a hole there and the links
which surround
the plaquette lie on the boundary  of the surface. So one can try to
describe the
boundary in the Chern-Simons theory using the algebra
(\ref{LC1},\ref{LC2}).
The theory living on the boundary is the chiral WZNW model. It is not
topological
and we cannot hope to describe it  in an adequate way using our rough
lattice
approximation. On the other hand, if one increases the number of
lattice sites so that
the distance between them becomes smaller and smaller,  the lattice
exchange
algebra   (\ref{LC1},\ref{LC2}) admits a nice continuous limit. Under
certain
assumptions it is possible to prove that this continuous limit
coincides with
the Kac-Moody algebra which governs the WZNW model assigned to the
boundary.
It was the reason to introduce the lattice exchange algebra
(\ref{LC1},\ref{LC2})
as lattice current algebra in \cite{AFS}.

So, for an appropriate choice of ciliation the graph connection
exchange
algebra includes lattice Kac-Moody algebras
for particular plaquettes as its subalgebras. It is one extra
check of consistency of our lattice model.

We have described the basic structures that we are going to
investigate
in this paper. Let us remark that the lattice simulations of the
Chern-Simons
theory and of the moduli space of flat connections are expected
to give exact results because the Poisson structure and the phase
space may be
reproduced exactly on the semiclassical level. The quantum theory on
the graph appears to be a lattice gauge theory associated
to the quantum group. This theory enjoys the quantum gauge symmetry
and this is the main difference between our model and  the model
\cite{Bou} where the relations (\ref{Ex3+}-\ref{Ex3-})
are replaced by the commutative relation of the type (\ref{Ex1}).
It is remarkable that the quantization of the Chern-Simons theory
leads
to quadratic algebra which uses R-matrices as structure constants.
It makes the theory efficiently finite dimensional and
this is the reason to call this approach {\em combinatorial
quantization} of the Chern-Simons model.

Now we change the language to a more mathematical one and turn to the
systematical
treatment of the algebra of observables of the Hamiltonian
Chern-Simons theory.

\section{The algebra $\G$ of gauge transformations}
\setcounter{equation}{0}

This section is devoted to a precise formulation of local gauge
symmetries on the graph or lattice. Gauge transformations $\xi$
will be assigned to the vertices of the graph. The algebra $\G$ of
all  gauge transformations comes equipped with  the structure of
a ribbon Hopf-*-algebra.

\subsection{The algebraic structure of $\G$}

To be specific, we
consider a graph $G$ formed by the edges and vertices
of a triangulation of a given oriented Riemann surface $\Sigma $.
For every oriented link $i$ of $G$ there is an oriented
link $-i$ which has opposite orientation. The set of
oriented links $i,-i,j,-j,k,-k,\dots$ will be denoted
by $L$. For elements in the set $S$ of vertices we use
the letters $x,y,z$.
We introduce the map $t:L \mapsto S$ such that
$t(i)=x$, if the oriented link $i$ points towards the vertex $x$.

We describe the local gauge symmetry by assigning a ribbon
Hopf-*-algebra $\G_x$ to every vertex $x \in S$. Ribbon Hopf-algebras
were introduced in \cite{ReTu}. Their definition is based on
the algebraic structure of quasitriangular Hopf-*-algebras, so
that the algebras $\G_x$ come equipped with a
co-unit $\e_x$, a co-product $\D_x$, an antipode $\S_x$
and an $R$-matrix $R_x$. While we assume the reader
to be familiar with the defining properties of a quasitriangular
Hopf algebra, we want to make some more detailed remarks on the
$*$-operation. In a Hopf-*-algebra co-product, co-unit and antipode
have to be consistent with the conjugation $*$. In detail this
implies that
$\e_x$ and $\D_x$ are *-homomorphisms, i.e.
$$ \e_x(\xi^*) = \overline{\e_x(\xi)}\ \ \ \ , \ \ \
\D_x (\xi^*) = \D_x (\xi)^* \ \ .$$
Since $\D_x(\xi)$ is an element of $\G_x\o \G_x$, the second
equation requires an action of $*$ on $\G_x\o \G_x$.
This action is not unique. One can either define
$(\xi \o \eta)^* = \xi^* \o \eta^*$ or (cp. \cite{MSIII})
\be
(\xi \o \eta)^* = \eta^* \o \xi^*  \ \ . \label{conj}
\ee
Throughout this paper we will consider the second case
(\ref{conj}). The main
reason is that this type of $*$-operation appears in many
interesting examples, e.g. in $U_q(sl_2), q^p=1$. Readers interested
in
the first case can easily rewrite everything below. The construction
of a scalar product on the space of physical states simplifies
dramatically.

It is consistent to demand that the antipode $\S_x$ is a
*-anti-homomorphism \cite{Sche},
$$ \S_x(\xi^*) = \S_x(\xi)^* \ \ .$$
In a quasi-triangular Hopf-*-algebra, unitarity of the $R$-matrix
$R_x = \sum_\s r^1_{x\s} \o r^2_{x\s}$,
\be R_x^* = \sum_\s r^{2*}_{x\s} \o r^{1*}_{x\s} = R_x^{-1}
\ee
is assumed to hold. Again these  properties can easily be
checked in the example $U_q(sl_2), q^p=1$.

Now let us proceed towards a description of ribbon Hopf-*-algebras
\cite{ReTu}. Given the $R_x$-element, we build $ u_x \in \G_x$
from its components,
$$ u_x = \sum_\s \S_x (r^2_{x\s}) r^1_{x\s}\ \ . $$
The standard properties of the element $u_x$ are
\ba
u_x \S_x^{-1}(\xi) = \S_x (\xi) u_x \ \ &,& \ \
u_x^* = u_x^{-1} \ \  \label{u}  \\[2mm]
\D_x(u_x) = (u_x \o u_x) (R_x'R_x)^{-1} & = &
(R_x'R_x)^{-1} (u_x\o u_x)\ \ .   \label{uR}
\ea
Moreover, the combination $u_x \S_x(u_x)$ is in the center of
$\G_x$. To obtain a ribbon Hopf-*-algebra we postulate
the existence of a central ``square root'' $v_x$ of this element
which is supposed to obey
\ba
v_x^2 = u_x \S_x(u_x) \ \ & , & \ \ \S_x (v_x) = v_x \ \ , \ \ \
\e_x(v_x) =1 \ \ ,\label{v}      \\[2mm]
v_x^* = v_x^{-1}\ \ & , & \ \
\D_x(v_x)  =   (R_x'R_x)^{-1} (v_x \o v_x) \label{eigRR}\ \ .
\ea
Such elements are known to exist for the quantized  universal
enveloping algebras of all simple Lie algebras \cite{ReTu}.

One could demand that all the algebras $\G_x$ are
isomorphic as Hopf algebras. But this is more than we
need. To prepare for a weaker statement let us recall
the notion of twist equivalence. $\G_x$ is said to be
{\em twist equivalent} to another ribbon Hopf-*-algebra
$\G_\ast$ with co-unit $\e_\ast$, co-product $\D_\ast$, antipode
$\S_\ast$, $R$-matrix $R_\ast$ and ribbon element $v_\ast$,
if there is a *-isomorphism $\iota_x:\G_x \mapsto \G_\ast$
such that
\ba
\e_x (\xi) = \e_\ast(\iota_x(\xi))  \ \ & , & \ \
\iota_x(\S_x(\xi)) = \S_\ast(\iota_x (\xi))  \ \ , \nn \\[2mm]
(\iota_x \o \iota_x)
(\D_x(\xi)) & = & F_x^{-1} \D_\ast (\iota_x (\xi))F_x \ \ ,\\[2mm]
(\iota_x \o \iota_x)(R_x) =  {F'}_x^{-1} R_\ast  F_x
\ \ \ &,& \ \ \ \iota_x (v_x) = v_\ast \nn\ea
holds for all $\xi \in \G_x$. Here $F_x \in \G_\ast \o \G_\ast$
is unitary, i.e. $F_x^* = F_x^{-1}$, and $F'_x$ denotes the
same element with exchanged components in the tensor product.
If we would restrict ourselves to  $F_x = e \o e$, we
would end up with the usual notion of isomorphic Hopf-*-algebras.
For the moment we assume that both co-products $\D_x,\D_\ast$
are co-associative. This amounts to a severe restriction on $F_x$.
However one can
check that $F_x = R^{-1}_\ast$ is related to a non-trivial
twist, which gives $(\iota_x \o \iota_x)(R_x) = {R'}_\ast^{-1}$.
Using this weak notion of equivalence of Hopf-*-algebras it
is natural to demand that {\em all the algebras $\G_x$ are twist
equivalent to the same ribbon Hopf-*-algebra $\G_\ast$}.
In other words we assume the algebras $\G_x$ of local
gauge transformations to be pairwise twist-equivalent.
Let us mention that the element $u_x$ introduced above
is independent of the twist in the sense that
$$ \iota_x( u_x) = u_\ast\ \ .  $$

The full gauge symmetry
$\G$ is obtained as a product over all local gauge symmetries $\G_x$,
$$\G \ = \ \bigotimes_{x \in S}\  \G_x\ \ .$$
The algebraic structure of the local symmetries induces
a co-product $\D$, a co-unit $\e$, an antipode $\S$ and a $R$-matrix
for the full gauge symmetry $\G$ such
that $\G$ becomes a quasitriangular Hopf-*-algebra in the sense
discussed above. Ribbon elements $v_x \in \G_x$ furnish a
ribbon element $v$ for $\G$.

\subsection{Representation theory of $\G$}

We start a discussion of the representation theory of $\G$ with some
general remarks. Given two representations $\t,\t'$ of a
Hopf-algebra $\G$, their {\em tensor product} $\t\bo \t'$ is defined
with the help of the co-product $\D$
$$
(\t \bo \t') (\xi) \equiv (\t \o \t') (\D(\xi))\ \ .
$$
The co-unit $\e$ is a one-dimensional representation of $\G$.
It is a trivial representation in the sense that
$(\e \bo \t)(\xi) = \t(\xi) = (\t \bo \e)(\xi)$ holds for
all $\xi \in \G$. A representation $\t$ on a Hilbert space $V$ is
called {\em unitary}, if $\t(\xi^*) = \t(\xi)^*$ for all $\xi \in
\G$. Note that the tensor product of two
unitary representations $\t,\t'$ is not unitary in general (provided
that we use the standard scalar product on the tensor product of
Hilbert spaces). Instead we have
$$ (\t \o \t')(\D'(\xi^*)) = ((\t \o \t')(\D(\xi))^* \ \ . $$
The (nonunitary) matrix $(\t \o \t')(R)$ furnishes an intertwiner
between the representations $(\t \o \t') \circ \D'$ and
$(\t \o \t') \circ \D$.

There are two natural ``contragradient'' representations
which come with the antipode $\S$. They are obtained as
\ba
(1) \ \ & & \ \ \tt (\xi) \equiv\ ^t\t(\S^{-1}(\xi))\ \ , \\
(2) \ \ & & \ \ \bt (\xi) \equiv\ ^t\t(\S(\xi)) \ \ ,
\ea
for all $\xi \in \G$. Here $\ ^t$ denotes the transpose
of matrices. The relations (\ref{u}) assert that $\tt$
and $\bt$ are equivalent but non-equal unless $u = e$.
Unitarity of $\t$ results in the unitarity of both
contragradient representations.
The tensor products $\t \bo \bt, \bt \bo \t$  contain
$\e$ as a subrepresentation (hence the name ``contragradient'').
These properties can be abstracted from the relations
\be
\sum_\s \bt_{ab} (\xi^1_\s) \t_{ac} (\xi^2_\s) = \e(\xi)
\d_{b,c} \ \ ,
\sum_\s \t_{cb} (\xi^1_\s) \bt_{ab} (\xi^2_\s) = \e(\xi)
\d_{a,c} \ \ ,
\label{contr} \ee
which are a direct consequence of the definition of the
antipode $\S$. Here $\xi^{i}_\s$ are defined via the decomposition
of
the coproduct:
\be
\Delta (\xi)= \sum_\s \xi^1_\s \o \xi^2_\s  \ \ .
\ee

Representations of the algebra $\G$ of gauge transformations
are obtained as families $ (\t_x )_{x \in S}$ of representations
of the symmetries $\G_x$. We are mainly interested in
those representations of $\G$ which come from the same representation
of $\G_\ast$. At this point let us assume that {\em $\G_\ast$ is
semisimple} and that every equivalence class $[J]$ of irreducible
representations of $\G_\ast$ contains a unitary representative
$\t_\ast^J$ with carrier space $V^J$.
For the moment, the most interesting examples of
gauge symmetries -- e.g. $U_q(sl_2), q^p = 1$ -- are ruled out by
this assumption. This will be revisited in section 7.
Tensor products $\t^I_\ast \bo \t^J_\ast$ can be decomposed
into irreducibles $\t^K_\ast$. This decomposition determines
the Clebsch-Gordon maps $C_\ast^a[IJ|K]: V^I \o V^J \mapsto V^K$,
\be
C^a_\ast [IJ|K] (\t^I_\ast \bo \t^J_\ast) (\xi) =
\t^K_\ast(\xi) C^a_\ast[IJ|K]\ \
{}.
\ee
The same representations $\t^K_\ast$ in general appears with some
multiplicity $N^{IJ}_K$. The superscript $a= 1, \dots, N^{IJ}_K$
keeps track of these subrepresentations. It is common to call
the numbers $N^{IJ}_K$ {\em fusion rules}. Normalization of
Clebsch Gordon maps is connected with an extra assumption.
It will be central for the positivity later.
Notice that the ribbon element $v_\ast$ is central so that the
evaluation with irreducible representations $\t_\ast^I$
gives complex numbers $v^I = \t_\ast^I(v_\ast)$.
We suppose that there exists a set of square
roots $\k_I, \k_I^2 = v^I, $ such that
\be
C^a_\ast[IJ|K] (\t^I_\ast \o \t^J_\ast) (R'_\ast)
C^b_\ast[IJ|L]^*  =
\delta_{a,b} \delta_{K,L} \frac{\k_I \k_J }{\k_K} \ \ .
\label{pos}
\ee
Here $R'_\ast = \sum r^2_{\ast\s} \o r^1_{\ast\s}$ .
Let us analyze this relation in more detail. As a consequence of
intertwining properties of the Clebsch Gordon maps and the $R$-element,
$\t^K_\ast(\xi)$ commutes with the left hand side of the equation.
So by Schurs' lemma, it is equal to the identity $e^K$ times some
complex factor $\omega_{ab} (IJ|K)$. After appropriate normalization,
$\omega_{ab}(IJ|K) = \d_{a,b} \omega(IJ|K)$ with a complex phase
$\omega(IJ|K)$. Next we exploit the $*$-operation
and relation (\ref{eigRR}) to find $\omega_{ab}(IJ|K)^2 =
v^I v^J/v^K$. This means that (\ref{pos}) can be ensured up to
a possible sign $\pm$. Here we assume that this sign is always
$+$. This assumption is met by the quantized  universal
enveloping algebras of all simple Lie algebras because they
are obtained as a deformation of a Hopf-algebra which clearly
satisfies (\ref{pos}).

We wish to combine the phases $\k_I$ into one element $\k_\ast$ in
the center of $\G_\ast$, i.e. by definition, $\k_\ast$ will denote
a central element
\be \k_\ast \in \G_\ast \ \ \ \mbox{ with } \ \ \t_\ast^J(\k_\ast) = \k_J
\label{k} \ \ . \ee
Such an element does exist and is unique. It has the property
$\k^*_\ast = \k^{-1}_\ast$.

The antipode $\S_\ast$ furnishes a conjugation
in the set of equivalence classes of irreducible representations.
We use $[\bar J]$ to denote the class conjugate to $[J]$. It is defined
such that the two equivalent contragradient representations $\tt_\ast^J$
$\bt_\ast^J$ of the representative $\t_\ast^J \in [J]$ are elements
in the conjugate class $[\bar J]$.

Let us finally mention that the trace of the element
$\S_\ast(u_\ast)v_\ast^{-1}$
in a given representation $\t^J$ computes the
``quantum dimension'' $d_J$ of the representation $\t^J$
\cite{ReTu}, i.e.
$$ d_J = \mbox{\it dim\/}_q(V^J)
\equiv \mbox{\it Tr\/}(\t^J(\S_\ast(u_\ast)v_\ast^{-1})) \ \ . $$

Representations and intertwiners of $\G_\ast$ are now transported
to the algebras $\G_x$. This is accomplished with the help
of isomorphisms $\iota_x$ and twist elements $F_x$.
\ba
 \t^I_x(\xi) &=& \t^I_\ast (\iota_x(\xi)) \ \ \mbox{ for all }\ \
 \xi \in \G_x \ \ ,    \nn \\[1mm]
 C^a_x[IJ|K] & = & C^a_\ast [IJ|K]
 (\t^I_\ast \o \t^J_\ast) (F^{-1}_x)\ \ .  \nn
\ea
The representations $\t^I_x$ act on the space $V^I_x = V^I$.
It is immediately checked that the new maps $C_x$ satisfy
the standard intertwining relations
\be
\t^K_x (\xi) C^a_x[IJ|K] = C^a_x [IJ|K] (\t^I_x \o \t^J_x)
(\D_x(\xi)) \ \ \mbox{ for all }  \ \  \xi \in \G_x \ \ .
\ee
Similar relations hold for the adjoint intertwiners
\ba
 C^a_x[IJ|K]^* & = &  (\t^I_\ast \o \t^J_\ast) (F'_x)
 C^a_\ast [IJ|K]^* \ \ ,     \nn \\[2mm]
 C^a_x[IJ|K]^* \t^K_x (\xi)
 & = & (\t^I_x \o \t^J_x) (\D'_x(\xi)) C^a_x [IJ|K]^*
 \ \ \mbox{ for all }  \ \  \xi \in \G_x \ \ .  \nn
\ea
Finally, the central element $\k_\ast \in \G_\ast$ introduced
in eq. (\ref{k}) is transported to central elements in $\G_x$
with the help of the formula $\k_x = \iota^{-1}_x (\k_\ast)$.

So far we have only described the representation theory
of the $\G_x$. Among all the representations of the total
algebra $\G$ of gauge transformations which can be built
from representations of the $\G_x$, we need only one
family $(\t^{I,i})_{i \in L}$ assigned to the links of the
graph. The representations $\t^{I,i}$ of $\G$ will later
describe the transformation properties of the basic
quantum variables  $U^I(i)$, i.e. of the parallel transporters
along the link $i$.
$$
\t^{I,i}(\xi) \equiv \left\{
\begin{array}{ll}
\t_y^I(\xi) & \mbox{if} \ \ \xi \in \G_{y}\\[1mm]
\ \bt_x^I(\xi) & \mbox{if} \ \ \xi \in \G_{x}\\[1mm]
\e_z (\xi) & \mbox{else}
\end{array} \right.
$$
for $x=t(-i), y=t(i)$. To decompose tensor products of
representations
$\t^{I,i}$, $\t^{J,i}$ assigned to the same link $i$, we use the
following intertwiners
\be \label{CGidef}
C^a[IJ|K]^i \equiv C^a_y[IJ|K] \o
\ ^t(C^a_x[IJ|K]^*)\ \ .
\ee
As usual, $\ ^t$ denotes the transpose. $C^a[IJ|K]^i$ is a map
from $(V_y^I \o V^J_y) \o (V^I_x \o V^J_x)$ to $V^K_y \o V^K_x$
which enjoys the intertwining property
\be    \label{CGi}
\t^{K,i} (\xi) C^a[IJ|K]^i = C^a [IJ|K]^i (\t^{I,i} \o \t^{J,i})
(\D(\xi)) \ \ \mbox{ for all }  \ \  \xi \in \G \ \ .
\ee

There are further relations between representations on the same
link, which involve both orientations $i,-i$.
In fact, there is an equivalence between the representations
$\t^{I,i}$ and $\tt^{I,-i}$. Let us describe this explicitly.
By rel. (\ref{u}), the element $\S_y(u_y)$ furnishes intertwiners
$\eta^I_y = \t^I_y (\S_y(u_y))$ with the property
$$  \eta^I_y \t^I_y(\S_y(\xi)) =
\t^{I}_y (\S^{-1}_y(\xi)) \eta^I_y\ \ . $$
{}From this equation one deduces that
\ba  \label{etai}
\eta^{I,i} \tt^{I,i}(\xi)  & = & \t^{I,-i}(\xi) \eta^{I,i}\ \ , \\
\mbox{ with } \ \ \eta^{I,i} & \equiv & e^I_x \o \ ^t\eta^I_y\ \ .
\nn
\ea
Here $e^J_x$ is the identity on $V^I_x = V^I$. A similar equivalence
appears between the representations $\t^{I,-i}$ and $\t^{\bar I,i}$.
This time the intertwiner is constructed from the Clebsch
Gordon maps. We introduce it according to
$$ \mu^{I,i} = n^I \ ^{t_2} C [I \bar I |0]^i (\eta^{I,i})^{-1} \ \
,
$$
where $\ ^{t_2}$ means
transpose only with respect to the second component and $n^I =
n^{\bar I}$
is a normalization determined by $\mu^{I,i} \mu^{\bar I, -i} = id$.
The element $\mu^{I,i}$ enjoys the intertwining property
\be  \label{mui}
\mu^{I,i} \t^{I,-i}(\xi)   =  \t^{\bar I,i}(\xi) \mu^{I,i}\ \ .
\ee

\section{Quantum group valued gauge fields}
\setcounter{equation}{0}

In this section we plan to introduce our basic lattice
algebra $\B$. It is an algebra generated by the quantum
lattice connections $U^I$ together with the quantum gauge
transformations $\xi \in \G$ discussed in the preceding
section. Relations between the elements $U^I$ and the
gauge transformations $\xi \in \G$ are determined by the
covariance properties of the quantum lattice connection.
All other relations among elements $U^I(i)$ are postulated
in the spirit of section 2.

\subsection{Definition of the lattice algebra $\B$}

To define the lattice algebra
$\B$ we have to introduce some extra structure on the
graph $G$. The orientation of the Riemann surface
$\Sigma$ determines a canonical cyclic order
in the set  $L_x = \{ i\in L : t(i)= x \}$ of
links incident to the vertex $x$. Writing the relations in $\B$
we are forced to
specify a linear order within $L_x$.
To this end one considers ciliated graphs
$G_{cil}$. A ciliated graph can be represented by picturing the
underlying graph together with a small cilium $c_x$ at each vertex.
For $i,j \in L_x $ we write $i \leq j$, if $(c_x,i,j)$
appear in a clockwise order.

\begin{defn} {\em (Lattice algebra $\B$) }   \label{algB}
The associative algebra $\B = \B (G_{cil})$
is generated by elements $U^I_{\a}(i)= U^I_{a_1a_2}(i)$,
$i\in L, \a = 1, \dots,$dim\/$(\t^{I,i})$,
and the elements of $\G$ such that
\begin{enumerate}
 \item the unit element $e$ of $\G$ acts as a
  unit element of $\B $, i.e.
  $U^I_\a(i)e=U^I_\a(i) = eU^I_\a(i)$.
 \item the tuples $(U^I_\a(i))$ transform covariantly according
  to the
  representation $\t^{I,i}$
  \be \xi U^I_\a(i) = U^I_\b(i) (\t^{I,i}_{\b\a}\o id)
  (\Delta (\xi )) \ \ \mbox{ for all}
  \ \ \ \xi \in \G  \ . \label{Ucov} \ee
 \item ``functoriality'' holds on the links
  \ba
  \label{OPE}
  U_\a^I(i) U_\b^J(i) & = & \sum U^K_\c (i) C^a\CG{I}{J}{K}
  {\a}{\b}{\c}^i  \ \ , \\
  \label{invers}
  U_{ab}^I(i) U_{cb}^I (-i) = \delta_{a,c} \ \ & , &
  U^I_{ba}(-i) U^I_{bc}(i) = \delta_{a,c}\ \ .
  \ea
  Here $C[..]$ are matrix elements of the Clebsch Gordon
  intertwiners (\ref{CGidef}) introduced in the last section.
 \item elements $U^I_\a(i)$ satisfy braid relations
  \be U^I_\a(i) U^J_\b(j) =  U^J_\c(j) U^I_\d(i)
  (\t^{I,i}_{\d\a} \o \t^{J,j}_{\c\b})(R)\ \ . \label{braid} \ee
  for $i\leq j$ or if i,j have no common endpoints.
\end{enumerate}
\end{defn}

This definition is rather central and requires some
thoughtful discussion. Intuitively, we prefer to think about
the generators $U^I_{a_1a_2}(i)$ as elements of a matrix.
Nevertheless proofs often simplify if we regard them
as vectors in a {\em dim\/}$(\t^{I,i})$-dimensional vector
space. Whenever we adopt the second point of view, we
use the multiindex $\a$ instead of its components $a_1,a_2$.

The covariance
relation in $2.$ can be written in a more explicit form if we
insert the expansion $\D(\xi) = \sum \xi^1_\s \o \xi^2_\s$.
$$ \xi U_\a^I (i)  = U^I_\b (i) \t^{I,i}_{\b\a}
 (\xi^1_\s) \xi^2_\s \ \ .$$
This tells us how to shift elements $\xi \in \G$ through
factors $U^I_\a(i)$ from left to right.
We note a simple consequence of
this fact.

\begin{prop}
Every element of $\B$ is a complex linear combination of elements
of the form
$$
U^{I_1}_{\a_1} (i_1) \dots U^{I_n}_{\a_n} (i_n) \xi \ \ \mbox{ with
}
\ \
n \geq 0, \xi \in \G    \ \ .
$$
\end{prop}

The relations (\ref{Ucov}) appear as a special case of a more general
notion of covariance.

\begin{defn}
{\em ($\G$ (right-) covariance)}
Let $\t = (\t_{\a\b})_{\a,\b \in I} $ be a representation
matrix of a n-dimensional representation of $\G$. An n-tuple
$F = (F_\a)_{\a \in I}, F_\a \in \B,$ is said to transform (right-)
covariantly according to the representation $\t$ of $\G$ if
\be
\xi F_\a = F_\b (\t_{\b\a} \o id)(\D(\xi))       \label{gcov}
\ee
for all $\xi \in \G$. $F \in \B$ is called $\G$-invariant if it
transforms according to the trivial representation $\e$ of $\G$, or
equivalently, if
\be
\xi F = F \xi
\ee
for all $\xi \in \G$.
\end{defn}

Indeed this is an appropriate notion of covariance. Assume for a
moment
that $\xi$ is an element of a unitary group rather than a
general Hopf algebra.
Then the co-product and the $*$-operation act according to $\D(\xi)
=
\xi \o \xi$ and $\xi^* = \xi^{-1}$. So the covariance relation
(\ref{gcov})
simplifies to $\xi F_\a \xi^* = F_\b \t_{\b\a} (\xi)$.

After this preparation we see that the covariance (\ref{Ucov})
can be regarded as a quantum version of the classical relation
(\ref{Latgauge}). The latter means that
the variable $U^I_{a_1a_2}(i)$ transforms
covariantly according to the representation $\t^I_y$ in the second
index
while it transforms according to the representation $\ ^t\t_x^I\circ
\S_x$
in the first index (if $i$ points from $x$ to $y$). This is
encoded in the definition of $\t^{I,i}$.

We will often have to move elements $\xi \in G$ from right to left.
According to the following proposition, this is always possible.

\begin{prop}  {\em (left covariance)}   \label{rlcov}
Suppose that the tuple $(F_\a), F_\a \in \B$ transforms covariantly
according to the representation $\t$ of $\G$. Then we have
\be
   F_\a \xi = (\tt_{\a\b} \o id) (\D(\xi)) F_\b
\ee
for all $\xi \in \G$. In other words, every right-covariant tuple
in $\B$ is also left-covariant.
\end{prop}

{\sc Proof:} We write the covariance relation for the components
$\xi^2_\s$ in the expansion of $\D(\xi)$.
$$ \xi^2_\s F_\a = F_\b \t_{\b\a} (\xi^{21}_{\s\t}) \xi^{22}_{\s\t}\
\ .
$$
Multiplication with $\t_{\a\c}(\S^{-1}(\xi_\s^1))$, summation
over $\s$ and the co-associativity of $\D$ lead to
\ba \xi^2_\s F_\a
\t_{\a\c}(\S^{-1}(\xi_\s^1)) & = &
F_\b \t_{\b\c} (\xi_{\s\t}^{12} \S^{-1} (\xi_{\s\t}^{11}))
\xi_\s^2  \nn \\
& = & F_\c \e(\xi^1_\s) \xi_\s^2 = F_\c \xi \nn \ \ .
\ea
The left hand side of this equation is equal to
$\tt_{\c\a}(\xi_\s^1) \xi_\s^2 F_\a$.
\\ \nm

This concludes our discussion of item $2.$ above. Let us
turn to functoriality next. At the end of the preceding
section we described a number of equivalences between
representations assigned to the link i. The relations
in $3.$ mean that all these equivalences reflect themselves
as equalities among the variables $U^I_\a(i)$.
While this explains the term ``functoriality'' it is
much more instructive to check that the postulated
relations are consistent with the covariance. This is done
by comparison of the definitions in $2.,3.$ with the
intertwining relations (\ref{CGi}) of
$C^a[IJ|K]^i$ and property (\ref{contr}).

Equations (\ref{OPE}) should be regarded as a kind of operator
product expansions. They can be written in a form which
comes close to the classical relations (\ref{CG}), if the
definition (\ref{CGidef}) of $C^a[IJ|K]^i$ is inserted.
The set of relations (\ref{invers}) reflect the behaviour
of $U^I(i)$ under $i \to -i$. In the formulation given in
$3.$ they look exactly like their counterparts (\ref{+-})
in section 2. In the quantum algebra $\B$ we would like to
substitute (\ref{invers}) by a new set of relations which
is manifestly covariant. Using the operator product
expansions (\ref{OPE}) one derives
\be    \label{imi}
   U^I_\a(-i)  =  U^{\bar I}_\b (i) \mu^{I,i}_{\b\a} \ \ .
\ee
(The element $\mu^{I,i}$ was defined in (\ref{mui})).
In fact, relations (\ref{OPE},\ref{imi}) are equivalent to
the pair (\ref{OPE},\ref{invers}) and thus furnish a
new definition of $\B$. The latter implies that
every product of elements $U^I_\a(i)$
and $U^I_\b(-i)$ is a complex linear combination of
$U^J_\b(i)$.

We can now proceed to the discussion of item $4.$. Of course braid
relations substitute for the commutation relations of classical
lattice connections. The braid relations between the components
of $U^I(i),U^J(j), i\leq j,$ are almost
uniquely determined by the consistency
with the transformation law and with the associativity of the
product in $\B$. Since $(\t^{I,i} \o \t^{J,j})(R)$ furnishes an
intertwiner
between the representations $\t^{I,i}
\bo \t^{J,j}$ and $\t^{J,j} \bo \t^{I,i}$,
both sides of the braid relations transform according to the
same representation $\t^{I,i} \bo \t^{J,j}$.
Consistency with the associativity
relies on the Yang Baxter equation for $R$. One should also notice
that
these braid relations  require the introduction of eyelashes.

Actually the braid relations in the definition of $\B$ are identical
to the corresponding relations in section 2. If $i,j$ have no common
endpoints then $(\t^{I,i} \o \t^{J,j})(R)$ is the identity matrix so
that
the corresponding variables $U^I(i),U^J(j)$ commute. Suppose next
that
the links $i,j$ point towards the same vertex $x$ while their second
endpoints are disjoint. Then $(\t^{I,i} \o \t^{J,j})(R)
= (\t^I_y \o \t^J_y)(R_y)$ and this matrix acts only on the second
component
of the indices $\c=(c_1,c_2),\d=(d_1,d_2)$. So we end up with the
relations
(\ref{Ex3+}) if $i < j $.
Finally we come to the case $i=j$, where the $R$-matrix in $4.$ picks
up
contributions from both endpoints of the link $i$.
More precisely $(\t^{I,i} \o \t^{J,j})(R)$ is equal
to the matrix $(\t^I_y \o \t^J_y)(R_y) (\ ^t\t^I_x \o \
^t\t^J_x)(R_x)$
acting on both components of the indices $\c,\d$. To see this one
uses that $(\S_x \o \S_x)(R_x) = R_x$. Consequently,
the braid relations (\ref{braid})  can be written in the
form of relation (\ref{Ex2}).

The braid relations spelled out in $4.$ do not determine the
commutation relations for arbitrary choice of the links $i,j$.
For example if $i,-j$ point towards the same endpoint $x$, the
commutation relations for $U^I(i),U^J(j)$ are not stated explicitly.
However they can be derived from the relations among
$U^I (i),U^{\bar J}(-j)$. The reason is that (\ref{imi})
provides a complex linear relation between $U^I(i)$ and
$U^{\bar I} (-i)$. As an example
we give the relations for $-i,-j$ if $i \leq j$.
\be
 U^I_\a(-i) U^J_\b (-j) = (\tt^{J,-j}_{\b\d}
 \o \tt^{I,-i}_{\a\d})(R') U^J_\d(-j) U^I_\c(-i) \ \ .   \label{--ex}
\ee
The reader is invited to verify this relation explicitly.
We arrive at the
rather important conclusion that any two
variables $U^I(i),U^J(j)$ can be
(braid-) commuted.

\begin{prop} \label{spanfin}
Suppose that $i_1,i_2, \dots i_n$ is a maximal ordered set of
oriented links with the property
that every link appears only once and only in one
orientation, i.e.  $\pm i_\nu \neq i_\mu$ for all $\nu \neq \mu$.
Then any element of $\B$ is a complex linear combination of elements
\be    \label{finform}
 U^{I_1}_{\a_1} (i_1) \dots
U^{I_n}_{\a_n} (i_n) \xi \ \ \mbox{ with } \ \
n \geq 0, \xi \in \G \ \ .
\ee
\end{prop}

The following proposition asserts that the functoriality
is consistent with the braid relations.

\begin{prop} {\em (braid relations for composite operators)}
\label{compbraid}
 Suppose that $F =(F_\a)$, $F'=(F'_\b)$ and $F''=(F''_\c)$
 transform covariantly according to representations $\tau $, $\tau'$
 and $\tau ''$ of $\G$.\\
 (i) Suppose that the braid relations
 \ba
  F_\a F'_\b & = &  F'_\mu F_ \nu
  (\tau_{\nu \a}\o \tau '_{\mu \b })(R ) \ \ , \nn \\
  F_\a F''_ \c & = &  F''_\mu F_ \rho
  (\tau_{\rho \a}\o \tau ''_{\mu \b })(R )\ \ ,  \nn
 \ea
 hold true. Then $F$ and $F'F''$ satisfy braid relations
 \be                         \nn
  F_\a F'_\b F''_\c  =
  F'_\mu F''_\nu F_\rho
  (\tau_{\rho  \a}\o (\tau '\bo \tau'')_{\mu \nu ,\b \c })(R ) \ \
.\nn
 \ee

\end{prop}

\noindent
The proof of this proposition is a standard application of the
quasi-triangularity relation of $R$ (cp.\cite{MSVI} for  details).

Before we finish our discussion on definition \ref{algB} we want to
remark
that an algebra similar to $\B$ was proposed by Boulatov \cite{Bou}.
In
his approach, variables assigned to different links commute.
We see that this is in general not consistent with the local
quantum symmetry of the model, i.e. by the  consistency with
the transformation law under local quantum symmetry transformations
one is forced to use braid relations instead of ordinary
commutation.

\subsection{The *-operation on $\B$}

We will obtain the observables of Chern Simons as a subalgebra of
$\B$. In quantum physics, observables come with a *-operation. This
*-operation will be a reminiscent of a *-operation in $\B$. The
construction of the latter is the main topic within this
subsection.

\begin{prop} {\em (anti-homomorphism $\th$) }    \label{sck}
There is a unique anti-homomorphism $\th:\B \mapsto \B$ with the
properties
\ba
\th(\xi) &=& \xi^* \ \ , \\
\th( U^I_\a (i)) &=& U^I_\c (-i) (\t^{I,-i}_{\c\b} \o id)
(R^{-1}) \eta^{I,i}_{\b\a} \label{kappa} \ \ .
\ea
In particular, $\th$ extends the $*$-operation on $\G \subset \B$.
\end{prop}

``Conjugations'' of this type were first proposed in \cite{MSIII}
(cp. also \cite{Sch1} for a simple example). If the $R$-matrix
would be trivial (as it is for group algebras), the action
of $\th$ would simplify to $\th(U_\a^I(i)) =
U_\a^I(-i)$. This is the familiar unitarity of the lattice
connection. The formula (\ref{kappa}) looks more convincing,
if we use the elements $R_x, R_y$ instead of $R$. One can  check that
\be
\th( U^I_{ad} (i)) = (\t_x^{I} \o id)_{ab}
 (R_x) U^I_{bc} (-i) (\t_y^{I} \o id)_{cd} (R_y^{-1}) \ \ .
\ee
We start the proof of the proposition with the following
lemma.

\begin{lemma}\ \ \  With $v_I = \t^I(v_\ast)$
the expression
(\ref{kappa}) for $\th$ can be rewritten according to
\be
\th(U^I_\a(i)) = v_I^{-2} \eta^{I,-i}_{\a\b}
 (\tt^{I,-i}_{\c\b} \o id)
 (R) U^I_\c (-i)\ \ .               \label{altkappa}
\ee
\end{lemma}

{\sc Proof:}  To prove this relation we apply the covariance
relation (proposition \ref{rlcov})
to move the $R$-matrix from the right to
the left and insert the definition of $\tt^{I,-i}$.
\ba
\th (U^I_\a(i)) &=& (\t^{I,-i}_{\c\b} \o
   \tt^{I,-i}_{\c\d} \o id)((id \o \D)(R^{-1}))
    U^I_\d(-i) \eta^{I,i}_{\b\a}\nn \\
      & = &
   (\tt^{I,-i}_{\b\c} \o \tt^{I,-i}_{\c\d}\o id)
   ((\S \o id \o id)(id \o \D)(R^{-1}))
    U^I_\d(-i) \eta^{I,i}_{\b\a} \nn \\
      & = &                 \nn
    (\tt^{I,-i}_{\b\d} \o id)((\S \o id)R^{-1}
    (\S(u^{-1}) \o e)) U^I_\d(-i) \eta^{I,i}_{\b\a}\ \ .
\ea
The last step uses the quasi-triangularity of $R$ and the definition
of $u$ (\ref{u}). Now  $(\S \o id) R^{-1} (\S(u^{-1}) \o e) =
(\S(u^{-1})\o e)
(\S^{-1} \o id) R^{-1} = (\S(u^{-1}) \o e) R$ and with the definition
(\ref{etai}) of $\eta^{I,i}$ this finally gives
the formula anticipated in the lemma.
\\[2mm]
{\sc Proof of proposition \ref{sck}: } Since the action of $\th$ is
specified on all generators of $\B$, uniqueness is obvious.
We have to show that the extension of $\th$ to $\B$
is consistent with the relations in $\B$. The simplest
part is the consistency with the covariance relations.
We apply $\th$ to the right hand side of
the covariance relation (definition \ref{algB}.2) and use
a series of intertwining relations.
\ba
& & \th (U^I_\b(i) (\t^{I,i}_{\b\a} \o id)\D(\xi)) \nn \\
 & = &
 v_I^{-2} (\t^{I,i}_{\a\b} \o id)\D'(\xi^*) \eta^{I,-i}_{\b\c}
 (\tt^{I,-i}_{\c\d} \o id)(R) U^I_\d (-i)  \nn\\
& = &
v_I^{-2}  \eta^{I,-i}_{\a\b}
 (\tt^{I,-i}_{\b\c} \o id)(R)
 (\tt^{I,-i}_{\c\d} \o id)\D(\xi^*) U^I_\d (-i)\nn  \\
& = &
v_I^{-2}  \eta^{I,-i}_{\a\b}
 (\tt^{I,-i}_{\b\c} \o id)(R) U^I_\c (-i)
  \xi^* = \th(\xi U^I_a(i)) \ \ .   \nn
\ea
The reader is invited to check this calculation carefully. The
consistency
with the covariance relations provides the main motivation for the
definition of $\th$. The factor
$$ (\t^{I,-i}_{\c\b} \o id)(R^{-1}) \eta^{I,i}_{\b\a}  $$
which appears in $\th (U^I_\a(i))$ is designed to match the
different transformation
laws of $\th (U^I_\a(i))$ and $U^I_\b(-i)$.

Let us turn to the braid relations next.
If $i,j$ have no common endpoints, the relations which result after
the applications  of $\th$ are obviously identical to the commutation
relations
for $U^I(-i),U^J(-j)$. So let us concentrate on the case $i \leq j$.
We have to check that
$$ \th (U^J_\b(j)) \th(U^I_\a(i)) =
(\t^{I,i}_{\a\d} \o \t^{J,j}_{\b\c})(R'^{-1})
\th(U^I_\d (i)) \th(U^J_\c (j))\ \ .
$$
To do this we insert the formula for $\th(U)$ given in the lemma
above.
After dividing by $v^{-2}_I \eta^{I,i}$  we obtain
\ba   & &       \nn
(\tt^{J,-j}_{\b\mu} \o id)(R) U^J_\mu(-j)
 (\tt^{I,-i}_{\a\nu} \o id)(R) U^I_\nu(-i)\\
& = &
(\tt^{I,-i}_{\a\d} \o \tt^{J,-j}_{\b\c})(R'^{-1})
(\tt^{I,-i}_{\d\rho} \o id)(R)
U^I_\rho (-i) (\tt^{J,-j}_{\c\s} \o id) (R) U^J_\s (-j)    \ \ .
  \nn
\ea
We apply the transformation law
(proposition \ref{rlcov}) to move all factors involving
$R$ to the left.
\ba   & &
(\tt^{I,-i}_{\a\nu} \o \tt^{J,-j}_{\b\mu} \o id)(R_{23} (id \o \D)R)
 U^J_\mu(-j)  U^I_\nu(-i) \nn\\
& = &
(\tt^{J,-j}_{\b\s} \o \tt^{I,-i}_{\a\rho} \o id)(R_{12}^{-1} R_{23}
(id \o \D) R) U^I_\rho (-i) U^J_\s (-j)     \ \ . \nn
\ea
The quasi-triangularity of $R$ helps to simplify the product of
$R$ matrices. More precisely, we apply
$$ (id \o \D)(R^{-1}) R_{23}^{-1} R_{12} R_{13} R_{23} R'_{12} =
R'_{12} $$
to end up with the formula (\ref{--ex}).
Consistency of $\th$ with the relations (\ref{OPE},
\ref{imi}) is left as an exercise.

One is tempted to guess that $\th$ gives a *-operation on $\B$ but
this is not quite true. In fact it turns out that $\th \circ \th$
is non-equal to the identity unless $R' = R^{-1}$. This is the
content of the following calculation.
\ba
\th \circ \th (U^I_\a(i)) & = & \th ( U^I_\c (-i)
(\t^{I,-i}_{\c\b} \o id )(R^{-1})
\eta^{I,i}_{\b\a} )\nn\\
& = & \eta^{I,i \ast}_{\a\b} (\t^{I,-i}_{\b\c} \o id)(R')
   \th (U^I_\c(-i))\nn \\
& = & \eta^{I,i\ast}_{\a\b} (\t^{I,-i}_{\b\c}
\o id)(R') v_I^{-2} \eta^{I,i}_{\c\d}
(\tt^{I,i}_{\d\nu} \o id)(R) U_\nu (i) \nn\\
& = &v_I^{-2} (\tt^{I,i}_{\a\nu} \o id)(R'R) U^I_\nu (i) \nn \\
& = & v U^I_\a (i) v^{-1} \ \ . \nn
\ea
The result gives a concrete idea how $\th$ can be improved
to obtain an involution. Recall that we have introduced
central elements $\k_x \in \G_x$ such that $\t^J_x(\k_x) =
v_J$. The set of $\k_x, x \in S,$ determines an element
$\k \in \G$ having the properties
$$ \k^2 = v \ \ \ \ \ ,  \ \ \ \ \k^* = \k^{-1}\ \ .  $$
Conjugation with $\k$ gives an automorphism of $\B$ which can
be used together with $\th$ to construct the desired *-operation
on $\B$. To formulate the result we define $\s_\k : \B \mapsto
\B$,
$$ \s_\k (B) = \k^{-1} B \k \ \ \mbox{ for all } \ \ B \in \B\ \ .$$

\begin{theo} {\em (*-operation on $\B$)} \label{kappa2}
The anti-automorphism $\s_\k \circ \th: \B \mapsto \B$ determines
a *-operation on $\B$, i.e. $(\s_k \circ \th)^2 = id$. We will
write $B^* \equiv (\s_k \circ \th) (B) $ for all $B \in \B$.
\end{theo}

\section{The *-algebra $\A$ of observables}
\setcounter{equation}{0}

We now come to the central part of this paper. The algebra $\A$ of
observables, i.e. invariant elements generated by the gauge field,
will be constructed. The *-operation on $\B$ can be restricted
to $\A$. Even though the gauge fields depend on the position of
cilia, the $*$-algebra
$\A$ is essentially independent and thus $\A (G_{cil}) = \A(G)$.
To avoid confusion about the term ``observables'' we should stress
that the observables of the Chern Simons theory are only obtained
after imposing the additional flatness conditions. This will be
discussed in a forthcoming publication. The true observable
algebras of Chern Simons will be identified as factor-algebras of
$\A$ and all statements we make about $\A$ in the following
-- in particular about the $*$-structure and positivity -- imply
corresponding results for Chern Simons observables.

\subsection{The definition of $\A$}

The elements $U^I_\a(i), i \in L $ generate a subalgebra of $\B$.
It will be denoted by $<U^I_\a(i)>$.

\begin{defn} {\em (algebra of observables)} The algebra $\A$ of
observables is the invariant subalgebra of $<U^I_\a(i)>$, i.e.
$$ \A \equiv \{ A \in <U^I_\a(i)> \subset \B | \xi A = A \xi \
\mbox{ for all } \ \xi \in \G\}\ \ . $$
\end{defn}

$\A$ is spanned by elements of the form
\be                                               \label{spanA}
C_{\a_1 \dots \a_n} U^{I_1}_{\a_1}(i_1)
\dots U^{I_n}_{\a_n}(i_n) \ \ \ n \geq 0   \ \ ,
\ee
where $C$ is supposed to possess the following intertwining property
$$ C (\tt^{I_1,i_1} \bo \dots \bo
\tt^{I_n,i_n})(\xi) = \e(\xi) C \ \ \mbox{ for all }
\ \ \xi \in \G \ \ .$$

Before we state our first result in this subsection we want to
introduce the following shorthand notation
\ba
 \D^{(1)} = \D &\ \ , &\ \ \ \ \ \   A^{(1)} = A
  \ \ ,  \nn \\   \label{defRn}
 \D^{(n+1)} &=& (id^{n} \o \D) (\D^{(n)})
 \ \ \ \mbox{ for all } \ \ n \geq 1\ \ , \\
 A^{(n+1)} &=& (id \o \D^{(n)}) (A)\, A^{(n)}  \ \ \
 \mbox{ for all }\ \ n \geq 1\ \ .         \nn
\ea
Here $A$ is an arbitrary element in $\G \o \G$.

\begin{theo} {\em (*-operation on $\A$)} \label{Astar}
The *-operation $\s_k \circ \th:\B \mapsto \B$ restricts to a
*-operation on the algebra $\A$ of observables.
\end{theo}

{\sc Proof:} We show that $\th$ maps all elements of the
form (\ref{spanA}) to elements in $\A$.
\ba
& & \th(C_{\a_1 \dots \a_n} U^{I_1}_{\a_1}(i_1)
\dots U^{I_n}_{\a_n}(i_n)) \nn \\
&=&
 \th (U^{I_n}_{\a_n}(i_n)) \dots \th(U^{I_1}_{\a_1}(i_1))
\overline{C_{\a_1 \dots \a_n}} \nn \\
& = &
 U^{I_n}_{\c_n}(-i_n) \dots U^{I_1}_{\c_1}(-i_1)
 (\t^{I,-i_n}_{\c_n\b_n} \o \dots
 \t^{I,-i_1}_{\c_1\b_1} \o id )((R^{-1})^{(n)})
C'_{\b_1 \dots \b_n} \nn\ \ .
\ea
Here we used the definition (\ref{defRn}) for $(R^{-1})^{(n)}$.
Factors
$\eta^{I,i}$ have  been absorbed
in the complex coefficients $C'_{\b_1\dots\b_n}$.
$C'$ has again a ``good'' intertwining property.
$$ (\t^{I_n,-i_n} \bo \dots \bo \t^{I_1,-i_1})(\xi)C' = C'\e(\xi)
\ \ \mbox{ for all }
\ \ \xi \in \G \ \ .$$
Since $(R^{-1})^{(n)}$ has $n+1$ components and we apply
representation
$\t^{I,-i}$ only to the first n components,
the above linear combination still has coefficients in $\G$.
However one can show (e.g. by drawing a picture),
\ba
& & (\t^{I_n,-i_n}_{\c_n\b_n} \o \dots
\t^{I_1,-i_1}_{\c_1\b_1}\o id )((R^{-1})^{(n)})
C'_{\b_1 \dots \b_n}  \nn \\
& = &
(\t^{I_n,-i_n}_{\c_n\b_n} \o \dots
\t^{I_1,-i_1}_{\c_1\b_1})((R^{-1})^{(n-1)})
C'_{\b_1 \dots \b_n} \ \ , \nn
\ea
so that the image of elements of the form  (\ref{spanA})
under $\th$ is indeed
an invariant element generated by the $U(i)'s$. Since $\k$
is in $\G$, it commutes with
all invariant elements in $\B$ and in particular with all the
observables. Hence $\s_k (A) = A$  for all $A \in \A$ to that
the assertion about $\s_k \circ \th$ follows from what we
said about $\th$.

\subsection{Independence of the eyelash}

The braid relations in $\B$ and hence $\B,*,\A$ depend on the
choice of the cilia at the vertices $x$. While this should not
disturb us as far as the ``unphysical'' algebra $\B$ is concerned,
we want the observable algebra $\A$ with $*$-operation $*$ to be
defined on the graph $G$ rather than on a ciliated graph $G_{cil}$.

\begin{prop} Suppose that $G_{cil}$ and $G_{cil'}$ are two ciliated
graphs which differ only by their ciliations. Then $\A(G_{cil})
\cong \A(G_{cil'})$ as $*$-algebras.
\end{prop}

{\sc Proof: }
Let us consider an elementary move
when the eyelash position changes
at one vertex $x \in S$ for one step.
This means that the smallest link incident
to $x$ (in the ciliated graph $G_{cil}$) becomes the largest
link incident to $x$ in $G_{cil'}$. This link will be
denoted by $i$. We agree to use $U^I_\a(j)$ for generators
in $\B( G_{cil} ) = \B$ and $\hat U^I_\a(j)$ for generators
of $\B( G_{cil'}) = \B' $. If $F \in \B$ we write
$\hat F$ to denote the corresponding element in $\B'$ where all
generators $U^I_\a(j)$ have been replaced by $\hat U^I_\a (j)$.
The only effect of the different position of eyelashes is
that the relations
$$ U^I_\a(i) U^J_\b(j) =  U^J_\c(j) U^I_\d(i)
  (\t^{I,i}_{\d\a} \o \t^{J,j}_{\c\b})(R_x)\ \  $$
which hold for all links $j \neq i$ on $G_{cil}$ incident to $x$
are substituted by
$$ \hat U^I_\a(i) \hat U^J_\b(j) =  \hat U^J_\c(j) \hat U^I_\d(i)
  (\t^{I,i}_{\d\a} \o \t^{J,j}_{\c\b})(R_x'^{-1})\ \  .$$
Observables in $\B$ are obtained as linear combinations of
$$ A = F_\mu U^I_\a (i) C_{\mu\a}\ \ .  $$
Here $F_\mu$ is a tuple of elements in $\B$ generated by
$U_\a^I(j), \pm j \neq i, $ and $F_\mu$ is supposed to transform
covariantly according to the representation $\t$ of $\G$.
$ C_{\mu\a} $ are complex numbers chosen in such a way that
$A$ is invariant. Let $A'$ be a second observable in
$\B$ which is written in the same form with a tuple $F'_\nu$
transforming according to the representation  $\t'$, i.e.
$A' = F'_\nu U^J_\b (i) C'_{\nu \b}$. The product $A A'$
defines coefficients $\tilde C_{\mu\nu\a\b}$ such that
$$ A A' = F_\mu F'_\nu U^I_\a (i)
U^J_\b (i) \tilde C_{\mu\nu\a\b} \ \ .$$
If we perform the same calculation in $\B'$ we obtain
$$ \hat A \hat A' =\hat  F_\mu \hat F'_\rho \hat U^I_\c (i)
\hat U^J_\b (i) (\t^{I,i}_{\c\a} \o
\t'_{\rho \nu})(R_x R'_x)^{-1} \tilde C_{\mu\nu\a\b} \ \ .$$
This basically follows from proposition \ref{compbraid} and the
influence of the different positions of the eyelash. Using the
intertwining properties of the coefficients $\tilde C$, the
equation $ (\S \o id)R = R^{-1}$,
relations (\ref{eigRR}) and functoriality on the link $i$
the product $\hat A \hat A'$ can be rewritten as
\ba
\hat A \hat A'  & = &
   \hat  F_\mu \hat F'_\nu \hat U^I_\c (i)
   \hat U^J_\d (i) (\t^{I,i}_{\c\a} \o
   \t^{J,j}_{\d \b})(R'_x R_x) \tilde C_{\mu\nu\a\b} \nn \\
    & = &
   \sum_K \hat  F_\mu \hat F'_\nu \hat U^K_\c (i)
   C^a \CG{I}{J}{K}{\a}{\b}{\c}^i
   \tilde C_{\mu\nu\a\b} \frac{v_I v_J }{v_K}\ \ . \nn
\ea
In the last line $v_I = \t_x^I(v_x)$ etc. . In other words, the
map $E: \A \mapsto \A'$ given by
$$ E(F_\mu U^I_\a (i) C_{\mu\a}) =
\hat F_\mu \hat U^I_\a (i) C_{\mu\a} v^{-1}_I  $$
is an automorphism, i.e. $E(AA') =  E(A) E(A') $.
Let us finally check that this automorphism
is compatible with the $*$-operation on observables.
\ba
E(A)^* & = &  (\hat F_\mu \hat U^I_\a (i) C_{\mu\a}  v^{-1}_I )^* \nn
\\
        & = & \overline{C_{\mu\a}} v_I
               \th(\hat U^{I}_\a (i)) \th(\hat F_\mu )  \nn \\
        & = &   \hat F'_\mu \hat U^{\bar I}_\a (i)
               \tilde C_{\mu\a}  v_I \ \ .  \nn
\ea
The last row is again meant to define the coefficients
$\tilde C$ and the tuple $\hat F'_\mu$. On the other hand we have
\ba
E(A^*) & = & E( F'_\rho U^{\bar I}_\b (i)
          (\t^{\bar I,i}_{\b\a} \o \t'_{\rho\mu})
           (R_xR_x') \tilde C_{\mu\a}) \nn\\
       & =& E ( F'_\nu U^{\bar I}_\a (i)
            \tilde C_{\mu\a}) v_I^2\nn\\
       & = &
       \hat F'_\nu \hat U^{\bar I}_\a (i)
        \tilde  C_{\mu\a} v_I\nn\ \ .\nn
\ea
So we conclude that $E(A^*) = E(A)^*$, i.e. $E$ defines a
*-automorphism between the algebras $\A$ and $\A'$. Since
two arbitrary ciliations of the graph $G$ can be obtained
from each other by a series the elementary moves  considered
in this proof, we established the independence of the
eyelash. Let us add that the isomorphisms between algebras
constructed starting from the same graph with different ciliations
are canonical. They are completely defined by the pair which
consists of the graph and
the symmetry Hopf algebra.

\section{The regular representation}
\setcounter{equation}{0}

Let us finally construct the regular representations of the
lattice algebra $\B$ and the subalgebra $\A$ of observables.
Both will act by multiplication operators on a space $\F$.
Elements in $\F$ have an interpretation as ``functions''
on the (noncommutative) space of lattice connections.
As an algebra, $\F$  is generated by a set of
``coordinate functions'' $u^I_\a(i)$. The elements of the gauge
symmetry $\G$ act on $\F$ as generalized derivations.
Using the $*$-operation on $\B$ and a generalized
multidimensional Haar measure we will be able to define
a scalar product on $\F$.

\subsection{The definition of $\F$}

In our present context it is obvious how to define the algebra
$\F$ of ``functions'' on the space of lattice connections.
So instead of giving a lengthy construction which works
also in the more general cases considered below, we present
an ad hoc definition for $\F$ and check that it carries
the announced representation of $\B$.

\begin{defn} {\em (Algebra $\F$) }   \label{algF}
The algebra $\F = \F(G_{cil})$ is an associative unital
algebra with unit $\Omega$. It is generated  by
elements $u^I_{\a}(i), i \in L, $ subject to the relations
\ba
  u_\a^I(i) \cdot u_\b^J(i) & = & \sum u^K_\c (i) C^a\CG{I}{J}{K}
  {\a}{\b}{\c}^i  \ \ ,\label{ubraid}  \\
   u^I_\a(-i) & = & u^{\bar I}_\b (i) \mu^{I,i}_{\b\a}
  \ \ , \label{uOPE}\\[1mm]
  u^I_\a(i)\cdot u^J_\b (j) & = & u^J_\d (j) \cdot u^I_\c (i)
  (\t^{I,i}_{\c\a} \o \t^{J,j}_{\d\b})(R) \ \ , \label{uinv}
\ea
for $i\leq j$ or if i,j have no common endpoints.
\end{defn}

We immediately recognize $\F$ to be our old algebra $<U^I_\a(i)>$.
However, this isomorphism is a mere coincidence. In the
more general framework of quasi-Hopf algebras, the analogue
of $\F$ turns out to be non-associative and consequently
cannot be isomorphic to any subalgebra of the associative algebra
$\B$. This remark might
seem too prospective, but it should at least explain why
we decided to give an independent definition of $\F$.

Because of their interpretation as functions in the
non-commutative coordinates $u = (u^I_\a(i))$, we will
often use symbols $\psi(u),\phi(u)$ for elements in $\F$.
A representation $\pi$ of $\B$ on $\F$ is defined
by the following relations
\ba
\pi (U^I_\a(i)) \psi(u) & = & u^I_\a(i) \cdot \psi(u)
\ \ ,    \label{piU}\\
\pi (\xi) (\psi_1 (u) \cdot \psi_2(u))  & = & \sum \pi(\xi^1_\s)
\psi_1(u)  \cdot \pi (\xi^2_\s)  \psi_2(u) \ \ ,
\label{pixi}  \\
\pi(\xi) u^I_\a (i) = u^I_\b (i) \t^{I,i}_{\b\a} (\xi) \ \ & , &
\pi(\xi) \Omega = \Omega \e(\xi)\ \      \nn
\ea
for all $\psi(u),\psi_1(u),\psi_2(u) \in \F$. The first equation
means that elements $U^I_\a(i)$ act as multiplication operators
on $\F$. The last two lines specify the action of the gauge
symmetry $\G$. Because of relation (\ref{pixi}) one says that
elements $\xi$ of the quantum symmetry act as {\em generalized
derivations} on the algebra $\F$.

We see that the operators $\pi(U^I_\a(i))$ generate $\F$ from the
``constant function'' $\Omega$. In particular $u^I_\a(i) =
\pi(U^I_\a(i)) \Omega$. Given an element $\psi(u) \in \F$,
its ``generator'' $\Psi(U) \in \B$
(i.e. $\psi(u) = \pi(\Psi(U))\Omega$) is nearly unique.
The only freedom in the choice of $\Psi(U)$ comes from
the possibility to multiply from the right by factors
$\xi \in \G, \e(\xi) = 1,$ without changing the
generated element $\psi(u) \in \F$.

A tuple $\psi_\a (u)$ of elements in
$\F$ is said to transform covariantly according to the
representation $\t$ of $\G$ if
$$ \pi(\xi) \psi_\a(u) = \psi_\b (u) \t_{\b\a} (\xi)\ \ . $$
$\phi(u) \in \F$ is invariant, if $\pi(\xi) \phi(u) =
\phi(u) \e(\xi)$. Invariant elements $\phi(u) \in \F$
generate a subalgebra $\F^{inv}$ of $\F$. This subalgebra
carries a representation of the algebra $\A$ of observables
(the restriction of $\pi$ to the algebra of invariants $\A$).
$\F^{inv}$ is the ``algebra of functions'' on the moduli
space of connections.

\subsection{A scalar product on $\F$}

It is our aim to construct a scalar product on $\F$, i.e.
for ``functions'' on the space of connections. The
procedure mimics the classical situation. The main ingredient
is a multidimensional Haar measure $\omega$, which allows to
compute integrals of arbitrary functions on the space of
connections. The scalar product of two functions
$\psi_i(u) = \Psi_i(U) \Omega , i = 1,2 $
is then obtained as $<\psi_2 (u)| \psi_1(u)> \equiv
\omega(\Psi_2(U)^* \Psi_1(U))$.

Instead of defining a functional directly on the algebra $\F$,
we prefer to work with a linear map  $\omega:\B \mapsto {\bf C}$.
The relation to the multidimensional Haar measure will be
apparent. By proposition \ref{spanfin}, a linear functional
on $\omega$ is specified when we prescribe the values
it has on elements of the form (\ref{finform}). In the
case of $\omega$ we want $\omega(e) = 1$ and
\be    \label{omega}
\omega( U^{I_1}_{\a_1} (i_1) \dots
U^{I_n}_{\a_n} (i_n) \xi )  = \e(\xi)
 \d_{I_1,0} \dots \d_{I_n,0} \ \ .
\ee
For this to be well defined it is essential that every link
appears only once and only in one orientation among the links
$i_\nu$. Some properties of $\omega$ are obvious. We state them
here without proof.
\ba
\omega(\xi F) & = & \omega (F) \e(\xi)  \label{Haarinv}\ \ , \\
\overline{\omega( F )} & = & \omega (F^*)   \label{Haark}
\ea
for all $\xi \in \G$, $F \in \B$.

The interpretation of $\omega$ as a multidimensional Haar
measure uses the correspondence between elements $\psi(u) \in
\F$ and their generators $\Psi(U)$. Since $\omega$ depends
on $\xi$ only through the value $\e(\xi)$, $\omega_h (\psi(u))
\equiv \omega (\Psi(U))$ does make sense. Relation
(\ref{Haarinv}) is the usual invariance $\omega_h(\pi(\xi)\psi(u))
= \omega_h(\psi(u)) \e(\xi)$ of the multidimensional
Haar measure.

Definition (\ref{omega}) has a fundamental drawback. Usually it
requires an enormous calculation to bring an arbitrary element
in $\B$ into the form (\ref{finform}). However there is a
recursive way to calculate $\omega$. Once all elements
assigned to a given link $i,-i$ are gathered, integration over
these variables can be performed. The formal expression is
\be            \label{defom}
\omega ( F  U^{I}_{\a}(i) G) \equiv
\omega(FG) \omega(U^{I}_{\a}(i)) = \omega(FG) \d_{I,0} \ \,
\ee
for all $F,G$ generated by $U^J_\b(j), j \neq i,-i$
and elements $\xi \in \G$.

Let us practise the calculation of the functional $\omega$
in a simple but fundamental example.

\begin{lemma} With the quantum dimension $d_J =
${\it Tr}$(\t^J(\S_\ast(u_\ast)v_\ast^{-1}))$
and $v_I = \t^I(v_\ast)$ we have
$$
\omega(\th(U_\a^I(i))U^I_\b(i)) = v^{-2}_I \omega(\eta^{I,-i}_{\a\c}
U^I_\c(i) U^I_\b(i)) =
\d_{\a,\b} \frac{v^{-1}_I}{d_J}\ \ .
$$
\end{lemma}

{\sc Proof:} The simplest proof for this formula makes
use of the invariance (\ref{Haarinv}) of $\omega$.
The latter can be reformulated into the following
intertwining property of the matrix $\Omega^{I,i}_{\a\b} =
\omega(\th(U_\a^I(i))U^I_\b(i))$.
$$
\t^{I,i}_{\a\b}(\xi) \Omega^{I,i}_{\b\c} = \Omega^{I,i}_{\a\b}
\t^{I,i}_{\b\c} (\xi) \ \ \mbox{ for all } \ \ \xi \in \G\ \ .
$$
Since $\t^{I,i}$ is irreducible, we obtain that $\Omega^{I,i}_{\a\b}=
\lambda^{I,i} \d_{\a,\b}$ . To calculate the complex number
$\lambda^{I,i} $ we multiply this equation with
$(\eta^{I,-i})_{\b\a}^{-1}$ and sum over $\a,\b$. The result is
\ba
\lambda^{I,i} Tr((\eta^{I,-i})^{-1}) & = & \sum
  (\eta^{I,-i})^{-1}_{\b\a} \omega(\th(U_\a^I(i))U^I_\b(i)) \nn \\
& = &  \sum v^{-2}_I \omega( U^I_\b (-i) U^I_\b(i))
= v^{-2}_I \delta_I \nn\ \ .
\ea
In this calculation we used the formula (\ref{altkappa}) for
$\th$,
the invariance (\ref{Haarinv}) of $\omega$ and the functoriality
on the link $i$ (\ref{invers}). $\d_I$ is the ordinary dimension
of the representation $\t^I_\ast$, i.e. $\d_I = ${\it dim\/}$(V^I)$.
Since $Tr((\eta^{I,-i})^{-1}) = \d_I
Tr(\t^I(\S_\ast(u^{-1}_\ast)))= v_I^{-1} \d_I d_I$ we
infer that $\lambda^{I,i} = v^{-1}_I / d_I $.

After this warm up we can address more complicated examples.
Recall that we plan to evaluate $\omega((\Psi_2(U))^* \Psi_1(U))$
for elements $\Psi_\nu (U) \in \B$. This motivates to
calculate
\be \label{allomega}
\omega( (U^{I_1}_{\a_1} (i_1) \dots
U^{I_n}_{\a_n} (i_n))^*  U^{I_1}_{\b_1} (i_1) \dots
U^{I_n}_{\b_n} (i_n) )
\ee
for a set of links $i_\nu$ which satisfies the assumption of
proposition \ref{spanfin}. We write the $*$-operation as
$\s_\k \circ \th$. The conjugation with $\k$ effects the value
of expression (\ref{allomega}) in a simple way. The $\k^{-1}$ to
the left gives a factor $\e(\k) = 1$ because of eq. (\ref{Haarinv}).
So there remains only one $\k$ in the middle. It can be removed
from the argument of $\omega$ by means of the covariance relation
(\ref{Ucov}) and invariance of the generalized Haar measure $\omega$.
After these manipulations $\s_\k$ is seen to contribute with
a factor $\k$ evaluated in the tensor product of representations
$\t^{I_1,i_1}$ through $\t^{I_n,i_n}$.
Now we are left with $\theta$. Since $\th$ is an anti-automorphism,
we find that all variables attached to the link $i_1,-i_1$ are
already gathered in the middle of the functional. So the
``integration'' can be performed and results in an expression
where variables on $i_2,-i_2$ appear together. This continues
until everything is reduced to the unit element $e$. Evaluation
on a single link is an application of the lemma. All one has
to care about are the factors involving the $R$-matrix which
come with the anti-automorphism $\th$.  The value of
(\ref{allomega}) is
\be  \label{allomegares}
 (\t^{I_1,i_1}_{\a_1\b_1} \o \dots \o
 \t^{I_n,i_n}_{\a_n \b_n}) ({R}^{(n-1)'}\D^{(n-1)}(\k)) \prod_\nu
 v^{-1}_{I_\nu} d_{I_\nu}^{-1}\ \ ,
\ee
where the $'$ means that one uses $\D'$ instead of $\D$
in definition (\ref{defRn})

\begin{theo} {\em (scalar product)}
Let $\Psi_\nu(U), \nu = 1,2$ be  elements in $\B$ which generate
the states $\psi_\nu(u)$ in the sense that
$\psi_\nu(u) = \pi(\Psi_\nu(U))\vac $.
The bilinear form
$$ <\psi_1(u)|\psi_2(u)>  \equiv  \omega(\Psi_1(U)^* \Psi_2(U))$$
defines a scalar product on $\F$ if and only if the
quantum dimensions $d_I$ satisfy $d_I > 0$ for all labels $I$
(and provided that the condition (\ref{pos}) is satisfied).
In particular, the assumption on $d_I$ guarantees that $\omega$
is a positive linear functional on $\A$.
\end{theo}

\noindent
{\bf Remark:} $\omega$ does not define a positive functional on
$\B$ or $<U^I_{\a}(i)> \subset \B$. For $\B$ this is due to the
fact that $\G \subset \B$. The properties of $\omega$ show
$\omega(\xi^* \xi) = \e(\xi^*) \e(\xi) = 0 $, whenever $\e(\xi)= 0$.
If $\G$ is not trivial, it has nonzero elements $\xi$ with
$\e(\xi) = 0$ and this obviously violates positivity.
On the other hand, $F^* \not \in <U^I_\a(i)>$ can happen, even
if $F \in <U^I_\a (i)>$. Consequently, the *-operation on $\B$ does
not restrict to $<U^I_\a(i)>$ so that there is no way to formulate
positivity of $\omega$ on $<U^I_\a(i)>$.

{\sc Proof of the theorem:} We remarked before that $< | >$ is well
defined. It is linear in the second argument and anti-linear in
the first. The property
$$ \overline{ <\psi_1(u)|\psi_2(u)>}  = <\psi_2(u)|\psi_1(u)> $$
holds due to relation (\ref{Haark}). Positivity is a consequence
of the formula (\ref{allomegares}) and our assumption (\ref{pos}).
If $C^\b_{\b_1 \dots \b_n}$ has the intertwining property
$$
 (\t^{I_1,i_1}_{\a_1\b_1} \bo \dots \bo
 \t^{I_n,i_n}_{\a_n \b_n}) (\xi) C^\b_{\b_1 \dots \b_n}
  = \t_{\b\a}(\xi) C^\a_{\a_1 \dots \a_n}   \ \ ,
$$
with $\t$ being some irreducible representation of $\G$, relation
(\ref{pos}) implies that  (no summation over $\a$)
$$
 \overline{C^\a_{\a_1 \dots \a_n}}  (\t^{I_1,i_1}_{\a_1\b_1} \o \dots \o
 \t^{I_n,i_n}_{\a_n \b_n}) ({R}^{(n-1)'}) C^\a_{\b_1 \dots \b_n}
 =
 \overline{C^\a_{\a_1 \dots \a_n}}
  C^\a_{\a_1 \dots \a_n} \k_\t^{-1} \prod_{\nu} v_{I_\nu}\ \ .
$$
Here we used  $\k_\t \equiv \t(\k)$ and $\k_J^2 = v_J$. We see that
the complex phases $v_{I_\nu}$ in the last expression cancel the phases
in (\ref{allomegares}), while the factor $\k_\t^{-1}$ cancels the
contribution from $\D^{(n-1)} (\k)$.
 With $F = U^{I_1}_{\a_1} (i_1) \dots
U^{I_n} _{\a_n} (i_n) C^\a_{\a_1 \dots \a_n}$ we obtain
$$
  \omega (F^* F) =
 \overline{C^\a_{\a_1 \dots \a_n}}
  C^\a_{\a_1 \dots \a_n} \prod_{\nu} \frac{1}{d_{I_\nu}} \ \ .
$$
This is positive, if the quantum dimensions are.

There is one important remark we have to make at this point.
Everything we did so far works for $U_q(\g )$ at generic
values of the deformation parameter $q$, since $U_q(\g )$
is semisimple in these cases. So it may seem that we just
quantized Chern-Simons for arbitrary (non-integer)
values of the level $k$.
However, the positivity of the scalar product is conditional
on the positivity of the quantum dimensions $d_J$. The latter
fails to hold for many representations of $U_q(\g )$, so that
we recover the usual quantization condition. In the next section
we will deal with the roots of unity. It is shown that
Chern Simons can be quantized only for primitive roots of
unity, i.e. for integer values of $k$.

\section{Generalization to quasi Hopf algebras}
\setcounter{equation}{0}

In this section we want to generalize our theory to cases in which
the
local symmetry $\G_x$ is a quasi-Hopf algebra. There are at least two
motivations to do this. When we discussed the twist equivalence
of the symmetry algebras $\G_x$ we saw that the co-associativity
of $\D_x$ led to a severe constraint on the possible twist. One
is tempted to remove this constraint and admit all possible
unitary twist elements $F_x$ without caring about co-associativity.
This is precisely what quasi-Hopf algebra are designed for.
We will see that our observable algebra
depends only on the ``twist class'' of the symmetry $\G_x$.
To understand the second motivation we recall that the assumption
about semisimplicity does not apply to the most interesting cases,
e.g. $U_q(sl_2), q^p=1$. However, working with the semisimple
``truncated'' symmetry algebra $U^T_q(sl_2)$ introduced in
\cite{MSIII}, we can bypass this problem.
By construction, the representation theory of $U^T_q(sl_2)$
coincides with the ``physical'' part of the representation
theory of $U_q(sl_2)$. There is a price we have to pay for this:
$U^T_q(sl_2)$ is no longer a Hopf algebra but only a
(weak) quasi-Hopf algebra.

\subsection{Short reminder on quasi-Hopf algebras}

At this point we want to recall some of the defining features
of (weak) quasi-Hopf algebras. Quasi-Hopf algebras have been
introduced in \cite{Dri2}. The axioms stated there can be
weakened to allow for ``truncation'' in the tensor product
of representations. The resulting structures were called
``weak quasi Hopf algebras'' \cite{MSIII}.

In comparison to Drinfel'd, we want to admit that $\D(e) \neq
e \o e$, where $e$ is the unit element in an algebra $\G$.
It still follows from the homomorphism property of $\D$,
$\D(\xi \eta) = \D(\xi) \D(\eta)$,  that $\D(e)$ is a
projector $P$ in $\G \o \G$ and that this projector commutes with
$\D(\xi)$ for all $\xi \in \G$.
Consequently, the linear map $(\t \o \t')(P)$ projects
onto a subrepresentation of $\t \bo \t'$. If $(\t \o \t')(P)
\neq id$, the tensor product $\t \bo \t'$ is said to be
truncated.

The generalization of Drinfel'ds axioms to the ``weak'' case
is almost straightforward. As in \cite{Dri2} we demand that
an element $\varphi \in \G \o \G \o \G$ is given which implements
(weak) quasi--co-associativity of the coproduct,
\be
\varphi (\Delta \o id ) \Delta (\xi ) =
(id \o \Delta )\Delta (\xi ) \varphi \mbox{ \ \ for all \ }
\xi \in \G . \label{coassco}
\ee
Because of the truncation, this element $\vp$ cannot be invertible
in general. So invertibility is substituted by a weaker assumption
on the  existence of a
quasi-inverse, still denoted by $\varphi^{-1}$, such that
\begin{eqnarray}
\varphi \varphi^{-1} = (id \o \Delta) \Delta (e) & ,  &
\varphi^{-1} \varphi = (\Delta \o id) \Delta (e) , \label{PMP} \\
(id \o \epsilon \o id) (\varphi ) &=& \Delta (e) \  . \nn
\label{epsphi}
\end{eqnarray}
The statement that $\vp^{-1}$ is a quasi-inverse of $\vp $ means that
$\vp \vp^{-1} \vp = \vp $. Evaluated with the representations
$\t,\t',\t''$, the re-associator $\vp$ furnishes an intertwiner
between the representations $(\t \bo \t')\bo \t''$ and
$\t \bo (\t' \bo \t'')$. This means that the tensor product
of representations is associative up to equivalence.

Similarly we do not demand that the element
$R$ be invertible. Instead it should have a
quasi-inverse $R^{-1}$ such that
\be
RR^{-1} = \Delta ' (e) \ \ \ , \ \ \ R^{-1}R  = \Delta (e)\ .
\ee
This is sufficient to implement the equivalence between
representations $\t \bo \t'$ and $\t' \bo \t$.

Following Drinfeld we postulate several relations
between $\Delta$,$R$, and $\vp$.
\begin{eqnarray}
   (id \o id \o \Delta )(\varphi )
   (\Delta  \o id \o id )(\varphi )
   & = & (e\o \varphi )      (id \o \Delta \o id )(\varphi )
    (\varphi \o e ) \ ,  \nn \\ 
   (id \o \Delta )(R) &=& \varphi_{231}^{-1}
R_{13}\varphi_{213}R_{12}
  \varphi^{-1} \ ,
  \label{A.13}
  \\
   (\Delta \o id  )(R) &=& \varphi_{312}
R_{13}\varphi_{132}^{-1}R_{23}
  \varphi \ .  \nn
  \end{eqnarray}
We used the standard notation. If s is any permutation of 123 and
$\varphi = \sum \varphi_{\sigma}^1
\o \varphi_{\sigma}^2 \o \varphi_{\sigma}^3$ then
\be
\varphi_{s(1)s(2)s(3)} = \sum_{\sigma} \varphi_{\sigma}^
{s^{-1}(1)} \o \varphi_{\sigma}^{s^{-1}(2)} \o \varphi_{\sigma}
^{s^{-1}(3)} \ . \nn
\ee
Eqs. (\ref{A.13})) imply validity of quasi
Yang Baxter equations,
  \be
  R_{12}\varphi_{312}R_{13}\varphi^{-1}_{132}R_{23}\varphi
  = \varphi_{321}R_{23}\varphi^{-1}_{231} R_{13}\varphi_{213}R_{12}
  \ , \label{QYBE}
  \ee
and this guarantees that $R$ together with $\varphi$ determines
a representation of the braid group \cite{MSVI}.

It is assumed that $(id \o \e \o id)(\vp) = \D (e)$. Similar
relations for the action of the co-unit $\e$ on other components
of $\vp$ and the components of $R$ follow from this.

All relations for the antipode $\S$ can be copied from Drinfel'd.
It is supposed that there is an anti-automorphism $\S$ and two
elements $\a,\b \in \G$, such that
\be
\sum \S(\xi_\s^1) \a \xi^2_\s = \a \e(\xi) \ \ \ , \ \ \
\sum \xi^1_\s \b \S(\xi^2_\s) = \b \e(\xi) \ \ .
\ee
Moreover, the following relations are required to hold.
\be \sum \vp^1_\s \b \S(\vp^2_\s) \a \vp^3_\s = e\ \ , \ \
    \sum \S(\phi^1_\s) \a \phi^2_\s \b \S( \phi^3_s) = e\ \ .
\ee
Here $\phi = \vp^{-1} = \sum \phi^1_\s \o \phi^2_\s \o
\phi^3_\s $.

Everything we said about *-structures in section 3 remains true.
But we have to add two more requirements which describe the
behaviour of the elements $\vp$ and $\a, \b$ with respect to
conjugation. One can check that the  equations
\be  \vp^*  =  \sum \vp^{3*}_\s \o
     \vp^{2*}_\s \o \vp^{1*}_\s = \vp\ \ ,
   \ \   \a^*  =  \b \nn
\ee
are consistent with the other relations which involve
$\vp$ or $\a,\b$.

For Hopf-algebras it is well known that
$\D (\phi) = (\S \o \S) \D'(\S^{-1}(\phi))$. A generalization
of this fact was already noticed by Drinfel'd \cite{Dri2}.
To state his observation we introduce the following
notations.
\ba
 \c &=& \sum \S(U_\s) \a V_\s \o \S(T_\s) \a W_\s \nn \\[1mm]
\mbox{with} & &                          \nn
\sum T_\s \o U_\s \o V_\s \o W_\s =
(\vp \o e)(\D \o id \o id)(\vp^{-1})\ \ , \\[2mm]
 f & = & \sum (\S \o \S)(\D'(\phi^1_p))  \c   \D (\phi^2_p \b
 \S(\phi^3_p))\ \ , \label{fel} \\[1mm]
\mbox{with} & & \phi = \vp^{-1} = \sum \xi^1_p \o
 \xi^2_p \o \xi^3_p\nn \ \ .
\ea
Drinfel'd proved in \cite{Dri2} that the element $f$ satisfies
\ba
 f \D(\xi) f^{-1}  & = &
 (\S \o \S) \D'(\S^{-1}(\xi)) \
 \ \mbox{ for all } \ \ \xi \in \G\ \ ,  \label{fint}  \\[1mm]
 \c & = & f \D (\a) \ \ . \nn
\ea
The first equation asserts that $f$ ``intertwines''
between the co-product $\D$ and the combination of $\D$ and
$\S$ on the right hand side.

Nontriviality of $\vp, \a$ effects the
expression for the element $u \in \G$. The relations
(\ref{u}) now hold for (cp. \cite{AlCo})
\be \label{qHu}
u = \sum \S (\phi^2_\s \beta \S(\phi^3_\s)) \S (r^2_\t ) \a r^1_\t
             \phi^1_\s  \ \ ,
\ee
with $\phi = \vp^{-1} = \sum \phi^1_\s \o \phi^2_\s \o \phi^3_\s$.
Altschuler and Coste also introduced the concept of
ribbon quasi-Hopf algebras. As before, $u\S (u)$ is central
and a ``square root'' $v$ with properties (\ref{v},\ref{eigRR})
it called ribbon element.

Examples of weak ribbon quasi-Hopf algebras
$\G$ are canonically associated with $U_q(sl_2)$ with
$q$ a root of unity.  As an algebra $\G \equiv U^T_q (sl_2) \equiv
U_q(sl_2)/\J$, where $\J$ is the ideal which is annihilated
by all the physical representations $\tau^I, 2I=0 \dots p-2$,
of $U_q(sl_2)$. $\U^T_q(sl_2)$ is semisimple, its
representations are fully reducible, and the irreducible
ones are precisely the physical representations of
$U_q(sl_2)$.
Let
\be  u(I,J)= \mbox{min}\{|I+J|, p-2-I-J \} \label{uIJ} \ee
and let $P_{IJ}$ be the projector on the physical subrepresentations
K, $|I-J| \leq K \leq u(I,J)$ of the tensor product
$\tau^I \bo_q \tau^J $ of $U_q(sl_2)$ representations.
There exists $P \in \G$ such that $P_{IJ} = (\tau^I \o \tau^J )(P)$.
The coproduct in $U^T_q(sl_2)$ is determined in terms of the
coproduct
$\Delta_q$ in $U_q(sl_2)$ as
\be
\Delta (\xi ) = P \Delta_q (\xi ) \ ,
\ee
hence $\Delta (e) = P \neq e\o e$. This coproduct specifies a tensor
product $\bo$ which is equal to the truncated tensor product of
physical $U_q(sl_2)$ representations. Thus
\be \tau^I\bo \tau^J  = \bigoplus_{|I-J|\leq K \leq u(IJ)} \tau^K \
{}.
\label{tensdec} \ee
   There exists an element
$\vp \in U^T_q(sl_2)^{\o_3}$
              such that $\vp^{IJK} = (\tau^I \o \tau^J \o \tau^K )
(\vp )$ imple
truncated
tensor products $\tau^I \bo ( \tau^J \bo \tau^K )$ and $(\tau^I \bo
\tau^J) \bo \tau^K $. The map
$\vp^{IJK}$ can be specified by its action on Clebsch Gordon
intertwiners, together with the condition
$\vp = (id \o \Delta )\Delta (e) \vp $, viz.
\be
C(IP|L) C(JK|P)_{23} \vp^{IJK}
= \sum_{Q}
\SJS{K}{J}{P}{I}{L}{Q}
C(IJ|Q) C(QK|L)_{12}\ ,
\label{defph}
\ee
where $C(..|.)$ denote the Clebsch Gordon maps and $\{ \}$ the
6J-symbols of $U_q(sl_2)$, $q^p=1,$ evaluated for physical labels
$I,J,K$. Summation over $Q$ is restricted to the physical
representations.

The $R$--element of
$U^T_q(sl_2) \o U^T_q(sl_2)$ is given in terms of the $R$--element
$R_q$ for $U_q(sl_2)$ by
\be
R= R_q \Delta (e) = \Delta ' (e) R_q \ ,
\ee
while antipode,* and co-unit are the same as in $U_q(sl_2)$.
One can show (cp. ref. \cite{MSIII}) that the defining properties of
a weak
quasitriangular quasi Hopf-*-algebra are satisfied. The *-operation
is of the type discussed above, i.e. $\D$ is a $*$-homomorphism
provided that $(\xi \o \eta)^* = \eta^* \o \xi^*$. The ribbon
element $v$ in $U_q(sl_2)$ survives the truncation and gives
a ribbon element in $U^T_q(sl_2)$.

The truncation procedure described here can be generalized
to other quantized enveloping algebras. We emphasize that the
assumption (\ref{pos}) holds for all truncated quantized
universal enveloping algebras associated with simple
Lie algebras.

\subsection{Results on quasi-quantum group gauge fields}

Our exposition will be restricted to the main results and
those parts which deviate from the above theory for
Hopf algebras.

A formulation of twist-equivalence involves the additional
relations
\ba
 (\iota_x \o \iota_x \o \iota_x)(\vp_x)
 & = & (e \o F^{-1}_x)
 (e \o \D_\ast)(F^{-1}_x)\  \vp_\ast \
 (\D_\ast \o e)(F_x) (F_x \o e) \ \ , \nn \\
 \iota_x (\alpha_x) & = &
 \sum \S_\ast(f^1_{x\s})\a_\ast f^2_{x\s} \nn
\ea
with $f^i_{x\s}$ defined through $F_x = \sum f^1_{x\s} \o
f^2_{x\s} $. Observe that the element $u_x$ introduced
in (\ref{qHu}) is independent of the twist, i.e. $\iota_x(u_x)=
u_\ast$. The same holds for the ribbon element $v_x$.

The discussion of gauge symmetry remains unchanged except
from some minor points in defining the intertwiners at the
end of section 3. First of all, while the definition of
Clebsch Gordon maps $C^a_x[IJ|K]$ at the sites $x \in S$
remain as before, the element $f$ constructed in eq.
(\ref{fel}) appears in the definition of the intertwiners
$$ C^a[IJ|K]^i \equiv C^a_y[IJ|K] \o \ ^t(
   (\t^I_x \o \t^J_x)(f'_x)  C^a_x[IJ|K]^*) $$
which are attached to the links $i \in L$. Of course we use
the element (\ref{qHu})
now to obtain the intertwiner $\eta^{I,i}$. The change in
$\mu^{I,i}$ is slightly more subtle. In the quasi-Hopf case
the right substitute is
$$ \mu^{I,i} = n^I\ ^{t_2}
 C [I \bar I |0]^i \t^{I,i}(\S^{-1}(\a))
(\eta^{I,i})^{-1} \ \ . $$
The last
change we have to mention concerns the definition of quantum
dimensions $d_I$. Their definition gets modified according
to (cp. \cite{AlCo})
$$ d_J
\equiv \mbox{\it Tr}(\t^J(\S_\ast(\b_\ast u_\ast)v_\ast^{-1}
\a_\ast ) \ \ . $$
Elements $\k_x \in \G_x$ and the corresponding element $\k \in
\G$ are defined as before.

The construction of the lattice algebra $\B$ is seriously effected
by the generalization.
To understand the major changes which occur when passing to quasi-
Hopf algebras, it is crucial to notice that due to the lack of
co-associativity, products of covariant elements are not
covariant in general. This motivates the definition of
``covariant products'' \cite{MSIII}.

Suppose that an algebra $\B$ contains
a quasi- Hopf algebra $\G$ as
subalgebra and that $(F_{\alpha })_{\alpha \in I},$
$(F'_{\beta })_{\beta \in I'}$ transform covariantly according to
representations $\tau $ and $\tau '$ of $\G$ with dimensions
$n$ and $n'$  (in the sense of definition \ref{gcov}).
Define the $\ti$-product of the components by
 \be
  (F_\a \ti F'_\b) =
   \sum_{\gamma\in I} \sum_{\delta\in I'}
   F_{\gamma }F'_{\delta }(\tau_{\gamma \alpha }\o
   \tau'_{\delta \beta }\o id )(\vp )\ \in \A  \ .
  \label{tiprod}
 \ee
Using the expansion $\vp =\sum \vp^1_{\s} \o \vp^2_{\s}
\o \vp^3_{\s}$ the defining
eq.(\ref{tiprod}) takes the form
 \be
  (F_\a \times F'_\b) = \sum_{\s}
  F_{\gamma }F'_{\delta } \tau_{\gamma \alpha }(\vp^1_{\s})
  \tau'_{\delta \beta }(\vp^2_{\s})\vp^3_{\s}  \ .
 \nn
 \ee
This exhibits the fact  that the $(F_\a \times F'_b)$ are
complex linear combinations of terms
$F_{\gamma }F'_{\delta }\vp^3_{\s}$ with
coefficients  $\vp^3_{\s}\in \G $.

\begin{prop} {\em (Properties of the $\times$-product)}
\label{propti}
 Let $(F_{\a}),(F'_{\b})$ be specified as above and suppose that the
 unit $e \in \G$ is a unit element in $\B$. Then
 the $\ti$-product (\ref{tiprod}) has the following properties.
 \begin{enumerate}
 \item
   Eq.(\ref{tiprod}) can be inverted
   to recover ordinary products
   from covariant ones, viz.
    \be F_{\a }F'_{ \b } =
     \sum ( F_\c\times F'_\d )(\tau_{\c \a }\o
     \tau'_{\d \b }\o id )(\vp^{-1} )\ .
    \label{covinv}
    \ee
 \item  The tuple $F\ti F' = ( F_\a \ti F'_\b)$
   transforms covariantly according to the tensor
   product representation $\tau \bo \tau'$ of $\G$. Hence we
   will often use the term {\em covariant product} instead of
   $\ti$-product.
  \item
   The $\ti$-product is not associative. But it is quasi--
   associative in the following sense. If  $F''=(F''_{\c })$
   transforms covariantly according to representations
   and $\tau ''$ of $\G $ and $F,F'$ as above, then
    \ba
     (i) \ \ \
     ((F_\a \ti  F'_\b)\ti F''_\c) & = &
     (F_\d\ti ( F'_\e \ti F''_{\kappa}))
     (\t_{\d \a } \o \t'_{\e \b}\o \t''_{\kappa \c})(\vp )
    \nn   \\
     (ii) \ \ \
     (F_\a\ti ( F'_\b \ti F''_\c))& = &
     ((F_\d\ti  F'_\e)\ti F''_\kappa)
     (\tau_{\d \a } \o \tau '_{\e \b}\o
     \tau ''_{\kappa \c})(\vp^{-1})\
     \ \ .
    \nn
    \ea
  \item
    If $G\in \B $ is $\G$-invariant then
   \be
    G\ti F_\a = GF_{\a }\ , \ F_a \ti G = F_{\a }G \ .
   \nn
   \ee
\end{enumerate}
\end{prop}

All items in this proposition follow from the properties of the
re-associator $\vp$. Proofs are spelled out in \cite{MSVI}.

Armed with the notion of covariant products
the definition of the lattice algebra $\B$ is straightforward.
We basically follow the rule to substitute all ordinary
products between generators $U^I_\a(i)$ by $\ti$- products.
So only functoriality and braid relations are concerned.
To implement functoriality on the link we divide by the
new relations
 \ba
  \label{qHOPE}
  U_\a^I(i) \ti U_\b^J(i) & = & \sum U^K_\c (i) C^a\CG{I}{J}{K}
  {\a}{\b}{\c}^i  \ \ , \\
  \label{qHinvers}
   U^I_\a(-i) & = & U^{\bar I}_\b (i) \mu^{I,i}_{\b\a} \ \ .
 \ea
while the braid relations become
 \be U^I_\a(i) \ti U^J_\b(j) =  U^J_\c(j) \ti U^I_\d(i)
  (\t^{I,i}_{\d\a} \o \t^{J,j}_{\c\b})(R)\ \ . \label{qHbraid}
 \ee
for $i\leq j$ or if i,j have no common endpoints. These relations
can be written in an alternative form involving again the ordinary
(associative) product in $\B$. For example the braid relations
can be formulated as
\be
  U_\a(i)  U_\b(j) =  U_\c (j) U_\d(i)
 (\t^i_{\d\a} \o \t^j_{\c\b} \o id )(\R) \ \ ,
\ee
where $\R = \vp_{213} R_{12} \vp^{-1}$ is the element which is
used to build up the representation of the braid group
in the quasi-Hopf case \cite{MSVI}.

The proposition \ref{spanfin} carries over to the more
general case but is is often more convenient to use another
set of linear generators built with the help of the
covariant product. By proposition \ref{propti}, elements
$$
 U^{I_n}_{\a_n} (i_n) \ti ( \dots \ti (
U^{I_2}_{\a_2} (i_2)\ti U^{I_1}_{a_1} (i_1)) \dots )
\xi \ \ \mbox{ with } \ \
n \geq 0, \xi \in \G \ \ .
$$
span $\B$. The brackets in this expression are necessary since
the $\ti$-product is not associative if $\vp$ is non-trivial.

Observe that the possibility to move elements $\xi \in \G$ was
an important ingredient in the Hopf-algebra case. However, the
proof of proposition \ref{rlcov} relied on the co-associativity.
Analogue formulas in context of quasi-Hopf algebras are more
complicated.

\begin{prop} {\em (right covariance)} Let
the element $w \in \G \o \G$ be
defined by $w \equiv \sum \vp^2_\s \S^{-1}
(\vp^1_\s \b) \o \vp^3_\s$ and
$m \in \G \o \G$ similarly as $m \equiv \sum \S (\phi^1_\s) \a
\phi^2_\s \o \phi^3_\s$ with the components $\phi^i_\s$ of
$\phi = \vp^{-1}$.
Suppose that the tuple $(F_\a), F_\a \in \B$ transforms
(right-) covariantly
according to the representation $\t$ of $\G$. Then the linear
combination
\be \label{rltrafo}
\pr{\a} F\equiv F_\b (\t_{\b\a}\o id)(w)
\ee
has the following properties.
\begin{enumerate}
\item
 $\pr{\a} F$ transforms left-covariantly according to the
representation
 $\tt$,
  \be
   \pr{\a} F \xi = (\tt_{\a\b} \o id) (\D(\xi))\pr{\b} F
  \ee
 for all $\xi \in \G$.
\item The transformation from right- to left-covariant elements
 can be inverted,
  \be
   F_\a = (\tt_{\a\b} \o id)(m) \pr{\b} F\ \ .
  \ee
\item The passage (\ref{rltrafo})
 from right- to left covariant elements is consistent with
 braid relations. Suppose that $F'_\b$ is a second
 right-covariant multiplet transforming
 covariantly according to the representation $\t'$ and that
 $F_\a, F'_\b$ satisfy braid relations
 $$  F_\a  F'_\b = F'_\d F_\c (\t_{\c\a} \o \t'_{\d\b} \o e)
  (\vp_{213} R_{12} \vp^{-1})\ \ .  $$
 Then the corresponding left covariant multiplets obey
 braid relations of the form
 \be
 \label{lcovbraid}
 \pr{\a} F \pr{\b} F' = (\tt_{\a\c} \o \tt'_{\b\d} \o e)
 (\vp_{213} R_{12} \vp^{-1}) \pr{\d} F' \pr{\c} F \ \ .
 \ee
\item On covariant products, the transformation (\ref{rltrafo})
 acts according to
 \be
 F_\a \ti F'_\b ((\t_{\a\c} \bo \t'_{\b\d}) \o e)(w))
 = (\tt'_{\d\b} \o \tt_{\c\a} \o e)(f_{12} \vp^{-1})
 \pr{\a} F \pr{\b} F'\ \ ,
 \ee
 where $f$ is the element (\ref{fel}). Note that the expression
 on the right hand side of the equation can be regarded as a
 linear combination of ``left covariant products''.
 \end{enumerate}
\end{prop}

We do not want to prove this proposition here. Details can be found
in
\cite{Sch3}. Especially the last two items are cumbersome. Note that
the
theorem establishes a complete symmetry between left and right.
It implies that it was just a matter of convenience to write all
the relations defining $\B$ in terms of the right covariant product.
They can all be rewritten in terms of a ``left covariant product''
and
even in a mixed form, where $\vp_x$ appears on the right and $\vp_y$
to the left of the product $U^I(i) U^J(j)$. In spite of this
left-right symmetry, there is now an algebraic difference between
components $U_\a^I(i)$ and the ``matrix-elements'' of a quasi-quantum
group valued gauge field. Matrix elements should be identified
with $\pr{a} U_b^I(i)$ and the latter differ from $U^I_{ab}(i)$.

Now we can proceed exactly as in the Hopf-algebra
case.  We define an anti-homomorphism $\th$ by
$\th(\xi) =  \xi^*$ and its action on left covariant elements,
\be
 \th(\pr{\a} U^I (i)) =  U^I_\c (-i)
   (\t^{I,-i}_{\c\b} \o id)(R^{-1}) \eta^{I,i}_{\b\a} \ \ .
\ee
On right covariant elements, $\th$ acts according to
\be
\th(U^I_\a(i)) = v_I^{-2} \eta^{I,-i}_{\a\b} (\tt^{I,-i}_{\c\b} \o id)
 (R) \pr{\c} U^I (-i)\ \ .
\ee
Although this second formula looks very familiar, it is
hard work to derive it from the definition of $\th$.
Consistency with the transformation law is a word by word repetition
of the above arguments. To check the compatibility with the braid
relations, we start from the equation (\ref{lcovbraid})
and apply $\th$. The rest is straightforward. Conjugation
with $\k \in \G$ improves this anti-automorphism $\th$ so that
it furnishes a $*$-operation $\s_\k \circ \th$ on $\B$.

The observable algebra $\A$ is obtained as an algebra of invariants
again. Let $<U^I_\a(i),\ti>$ denote the {\em linear
space} generated
the (right-) covariant product $\ti$. Observe that this space is
not closed under ordinary associative products in $\B$.
Within $<U^I_\a(i),\ti>$ only the quasi-associative multiplication
with the covariant product $\ti$ is possible. We define $\A$ to be
the
subspace of invariants, i.e.
\be
\A \equiv \{ A \in <U^I_\a(i),\ti> | \xi A = A \xi \mbox{ for all }
\ \ \xi \in \G \}\ \ .
\ee
Obviously $\A$ is closed under covariant multiplication and since
(by proposition \ref{propti}) the covariant product and the
associative
product coincide for invariants, $\A$ comes equipped with an
associative product.

\begin{theo} {\em (algebra $\A$ of observables)} Let $\G \equiv
\bigotimes \G_x$ be a ribbon quasi-Hopf algebra.  Then the
the associative algebra $\A$ of invariants in
the vector space $<U^I_\a(i), \ti >$ is a *-algebra with
$*$-operation $A^* = (\s_k \circ \th)(A)$. If the quantum
dimensions $d_I$
are positive and assumption (\ref{pos}) is satisfied,
the linear functional
$\omega: \B \mapsto {\bf C}$ defined by (\ref{omega})
restricts to a positive linear functional $\omega$ on $\A$, i.e.
$\omega(A^* A) \geq 0$ for all $A \in \A$. The algebra $\A$
is independent of the position of eyelashes.
\end{theo}

To construct the regular representation one can follow
a recipe given in \cite{MSVI}. We do not want to repeat the
individual steps here. Instead let us sketch the main
results. The analogue of $\F$ is {\em non-associative}  unital
algebra with unit $\Omega$. The generators $u^I_\a(i)$
of $\F$ can be braid commuted with (\ref{ubraid}) (i.e. the
braid relations of elements $u^I_\a(i)$ contain no $\vp$).
One can prove that eq. (\ref{uOPE},\ref{uinv}) continue to
hold. To define a representation $\pi$ of $\B$ on $\F$
one uses rel. (\ref{piU},\ref{pixi}). This provides us with
a notion of covariance and invariance in $\F$ in the same
way as before. Invariant elements in $\F$ form an
{\em associative} subalgebra $\F^{inv}$. Finally, the
functional $\omega$ can be used to define a scalar product
on $\F$ as this was done before.

The non-associativity of $\F$ is relatively harmless.
As one may guess by now, $\F$ turns out to be quasi-associative
in the sense that products with different positions of
brackets are linear combinations of each other. The
``re-association'' can be performed with the help of
$$
(\psi_\a \cdot \psi'_\b) \cdot \psi''_\c
   = \psi_\d \cdot (\psi'_\rho \cdot \psi''_\s )
   (\t_{\a\d} \o \t'_{\b\rho} \o \t''_{\c\s})(\vp) \ \ ,
$$
which hold whenever $\psi_\a,\psi'_\b,\psi''_c$
transform covariantly according to the representation
$\t,\t',\t''$ of $\G$. In the title of this subsection we
suggest the name ``quasi-quantum group'' for the algebra
generated by the $u^I_\a(i)$ on a given link $i$. Because of
the quasi-associativity of
the multiplication in $\F$, there is now a dramatic difference
between ``quasi-quantum groups'' and quasi-Hopf algebras.
At least for a special choice of the twists $F_x$ in the
endpoints, the algebra generated by the $u^I_\a(i)$ is dual
to the quasi-Hopf algebra $\G_\ast$. Since co-associativity
generically fails to hold in $\G_\ast$, this duality gives
another ``explanation'' why the  algebra of ``functions''
on the space of connections becomes quasi-associative.

As a corollary of the above theorem and the remarks we made
before, it is finally obtained that

\begin{coro}
Hamiltonian Chern Simons theory can be
quantized for all integer values of the level $k$ and for
every simple Lie algebra $\g $.
\end{coro}

This corollary needs some explanation because there are many senses
of the word ``quantization''. Here {\em to quantize} is treated in
the sense of deformation quantization.
Namely, one should construct an algebra of observables supplied with
a $*$-operation
and a positively defined trace functional. Consistence with the
Poisson brackets is required
in the classical limit. In this sense we obtained a quantization of
the algebra of lattice gauge
observables. In order to make the same statement about the whole
Hamiltonian Chern Simons
theory, we should impose the quantum flatness conditions in a way
consistent with the $*$-operation. We postpone this question to the
next paper.

\subsection{Twist equivalence}

We are now prepared to show that the algebra of observables actually
does only depend on the ``twist class'' of $\G_x$. Let us make this
statement precise. Suppose that $\hat \G_x$ is a second set of
quasitriangular quasi Hopf-*-algebras which are again twist
equivalent to the $\G_\ast$ but with  possibly different
twist elements $\hat F_x \in \G_\ast$. Then one can follow all the
steps described above to build a lattice algebra $\hat \B$ with
*-operation $*$, an algebra of invariants $\hat \A$
etc.. In general these structures will depend on the choice
of $\G_x$. The observable part of the theory, however,  does
not change.

\begin{theo} {\em (twist independence of $\A$)} Suppose that
$(\G_x)_{x\in S}$ and $(\hat \G_x)_{x \in S}$ are two families
of gauge symmetries and that $\G_x$ as well as $\hat \G_x$ are
twist equivalent to the quasitriangular quasi Hopf-*-algebra
$\G_\ast$ for all $x \in S$. Then there is a *-isomorphism
$i: \hat\A \mapsto \A$.
\end{theo}

{\sc Proof:}  We denote the ratio of the
twist elements by $h_x =  F^{-1}_x \hat F_x$. With the help of the
isomorphism $\iota_x$ this family is lifted to an element
$h \in \G \o \G$. There is an isomorphism $i: \hat\B \mapsto \B$
defined by $i(\xi) = \iota^{-1}_x \circ \hat \iota_x(\xi)$
for all $\xi \in \hat \G_x$ and
\be
i (\hat U^I_\a (i)) = U^{I}_\b (i)
(\t^{I,i}_{\b\a} \o id)(h)\ \ .  \ \
\ee
With all the experience we gained in prior calculations,
the homomorphism property is now almost obvious. So let
us directly proceed to the consistency with $*$. To prove
that $i$ is a *-homomorphism, one has to know, how
the element $w$ behaves under twists. The answer is
$$ w = \D(h_\s^2) h (i \o i)(\hat w)(\S^{-1}(h_\s^1) \o e) \ \ , \ \
h = \sum h^1_\s \o h^2_\s \ \ .$$
Let us agree from now on to identify elements in $\G$ and
$\hat \G$ with the help of the isomorphism $i$, so that,
for example, $\hat w$ can be regarded as an element in
$\hat \G \o \hat \G$ without the extra action of $(i \o i)$.
The twist dependence of $w$
yields the action of $i$ on left-covariant elements.
$$ i ( \pr{\a} {\hat U}^I (i)) =
(\tt^i_{\a\b} \o id) (h^{-1}) \pr{\b} U^I(i)\ \ .$$
The element $\eta$ is independent of the twist, and consequently
$\hat \eta^{I,i}$ and  $\eta^{I,i}$ can be identified in the
following calculation
\ba
  i(\hat \th(\pr{\a} \hat U^I(i)))
& = &  U^I_\c (-i) (\t^{I,-i}_{\c\b} \o id)(h \hat R^{-1})
       \eta^{I,i}_{\b\a} \nn \\
& = &  U^I_\c (-i) (\t^{I,-i}_{\c\b} \o id)(R^{-1} h')
       \eta^{I,i}_{\b\a}  \nn \\
& = &  \th(\pr{\d} U^I (i))
        (\tt^{I,i}_{\d\a} \o id)(h') \nn \\
& = &  \th((\tt^{I,i}_{\a\d} \o id)(h^{-1})
        \pr{\d} U^I (i)) =
        \th(i(\pr{\a} U^I(i))\ \ . \nn
\ea
So $i$ is compatible with the anti-automorphisms $\th, \hat \th$ and
hence with the $*$-operations on $\B, \hat \B$.
The proof of twist independence is complete if we can show that
$i$ restricts to a map between the observable algebras $\hat \A$
and $\A$. To see this we recall that observables are linear
combinations of elements
\be
 \hat U^{I_n}_{\a_n}(i_n) \hat \ti ( \dots
 \hat \ti (\hat U^{I_2}_{\a_2}(i_2)
 \hat \ti  \hat U^{I_1}_{\a_1}(i_1) )\dots )
 \hat C_{\a_n, \dots \a_2,\a_1} \ \ ,
\ee
where $\hat C$ is supposed to possess the usual intertwining
properties,
$$ (\t^{I_n,i_n} \hat{\bo} (\dots
\hat{\bo} (\t^{I_2,i_2} \hat{\bo} \t^{I_1,i_1})\dots ))
(\xi) \hat C = \hat C \e(\xi)$$
for all $\xi \in \hat G$. To evaluate $i$
on such elements one uses the following compatibility of $i$ with
covariant multiplication
\ba
  i(\hat U^I_\a (i) \hat\ti \hat U^J_\b (j))
   & = &
   U^I_\c(i) U^J_\s(j) (\t^{I,i}_{\c\a} \o \t^{J,j}_{\s\b} \o e)
   ((id \o \D)(h) h_{23} \hat \vp) \nn \\
   & = &
   U^I_\c(i) U^J_\s(j) (\t^{I,i}_{\c\a} \o \t^{J,j}_{\s\b} \o e)
   (\vp h_{12}(\hat \D \o id)(h)) \nn \\
   & = &
   U^I_\c(i) \ti U^J_\s(j) (\t^{I,i}_{\c\a} \o \t^{J,j}_{\s\b} \o id)
   ( h_{12}(\hat \D \o id)(h))\ \ . \nn
\ea
When we act on multiple $\ti$-products we obtain a similar expression
in which the argument of $ (\t^{I_n,i_n} \o \dots
\o \t^{I_1,i_1} \o id)$ splits into two factors.
The right factor essentially annihilates by the multiplication
with $\hat C$ while the left factor is trivial in the last
component and consequently gives rise to a complex linear
combination of
\be
  U^{I_n}_{\a_n}(i_n) \ti ( \dots  \ti ( U^{I_2}_{\a_2}(i_2)
  \ti   U^{I_1}_{\a_1}(i_1) )\dots )
  \hat C_{\a_n, \dots \a_2,\a_1} \ \ .
\ee
Since the image $i(G)$ of an invariant element $G \in \hat\B$
is automatically invariant, this proves the proposition.

\section{Outlook}
\setcounter{equation}{0}

In the first part of our work we discussed a new notion of
quantum group lattice gauge models. The algebra of observables
in such a model has been constructed. Our definitions were motivated
by the Fock-Rosly discussion of classical Chern Simons
theories. From our initial remarks on this topic it is
almost clear that there is a deep connection between
quantized Chern Simons theories and our lattice gauge
models. Let us sketch here the main  points that we
are going to  discuss in details in the second paper
devoted to the same subject.

The first important step towards the quantum  Chern-Simons theory
is to implement the flatness condition in
the lattice gauge theory. The idea is the following. Let us introduce
a monodromy matrix for each plaquette so that

\begin{equation}
M=U(i_{1})\dots U(i_{s}), \label{Mon}
\end{equation}
where the links $i_{1}\dots i_{s}$ surround the chosen
plaquette. Our purpose is to fix somehow the eigenvalues
of the matrix $M$ as we did in Section 2 (see formulae
(\ref{CExporb},\ref{QExporb})).  From the very beginning
it is not clear that there exists a quantum analogue of this
procedure.  It might happen that operators corresponding
to eigenvalues of $M$ do not commute and it would cause
serious problems. Fortunately,  these operators commute
with each other and moreover they belong to the center
of  the algebra of observables. So one can find a set of
central projectors which efficiently fix eigenvalues of
monodromy matrices for all elementary plaquettes.  One of
such projectors has been introduced in \cite{Bou} and called
quantum $\delta$-function. Having imposed the condition
of flatness we obtain the quantum algebra of functions
on the moduli space of flat connections (moduli algebra)
or, in other words, the algebra of observables of the
Hamiltonian  Chern-Simons theory. The moduli algebra
enjoys an important property. It does not depend on the graph
on the Riemann surface which we have started with \cite{FoRo}!
It is possible to prove that the moduli algebras constructed
starting from different graphs are canonically isomorphic to
each other. Thus, the algebra of observables is defined
by the symmetry algebra, Riemann surface with marked points
and by the set of representations of the symmetry algebra
assigned to the marked points. After the graph independence
of the construction is established we have a good chance
to prove that the lattice gauge model defines the same theory
as the continuous Chern Simons model.

The important question which we face at this stage is
how to construct the representation theory of the moduli algebra.
There are two different approaches to this problem. The first
of them  (let us call it algebraic) is based on the graph
independence of the moduli algebra. One can choose
the simplest possible graph and consider the moduli algebra
in some explicit coordinates given by the algebra of graph
connections on this chosen graph. For generic values of the
deformation parameter $q$ this method has been applied
in \cite{AA}. There the graph is chosen to be a bunch of
circles on a Riemann surface of genus $g$ with $n$ marked
points. The first $g$
circles represent $a-$ and $b$-cycles winding
around the handles of the surface and $n$ cycles which
surround the marked points.
Using this particular
graph one can describe the representation theory of the
moduli algebra quite efficiently. We shall fulfill this
program for $q$ being a root of unity.

The second approach to the representation theory
(let us call it geometric) is based on two simple observations.
First of all we already know the desirable answer.
The algebra of observables of the Hamiltonian Chern-Simons
theory must act in the Hilbert space of this theory. The latter is
isomorphic to the space of conformal blocks in the corresponding
WZNW model or, more technically, to the space of solutions of the
Knizhnik-Zamolodchikov equation satisfying certain conditions
\cite{Wit1,FrKi}.
The second observation concerns the nature of structure constants
which we use to define the algebra of graph connections. Basically
we use two objects,  $R$-matrix and the associator $\varphi$. Both of
them may be regarded as certain monodromy matrices for solutions
of the  Knizhnik-Zamolodchikov equation \cite{Dri2}. These
two facts enable us to represent the observable algebra directly in
the space of conformal blocks. Particular operators act as
combinations of monodromies of the Knizhnik-Zamolodchikov
equation. Let us mention that
the idea of the geometric approach is  essentially borrowed from
the combinatorial description of Vassiliev-Kontsevich knot
invariants \cite{Piun}, \cite{Cart}.

The geometric construction of the representation theory
of the moduli algebra will provide the representations which
are realized directly in the Hilbert space of the continuous
Chern Simons theory. This will be a final check of the conjecture
that the lattice gauge model presented in this paper indeed
coincides with the Hamiltonian Chern Simons theory.

The ideas which we have shortly described in this Section
will be considered in details in the forthcoming paper.

\section{Acknowledgments}

This work was initiated while two of us (A.A.,V.S.) stayed
at the Erwin Schr\"{o}\-dinger International Institute in Vienna.
We would like to thank the local organizers for
providing ideal working conditions and a pleasant
atmosphere. In the initial phase, V. V. Fock contributed
through intense discussions with one of the authors (A.A.).
Conversations with A. Beilinson, V.G. Drinfeld,
A. Jaffe, T. Kerler, A. Lesniewski,
G. Mack, F. Nill and K. Szlachanyi are gratefully acknowledged.

The work of V.S. was partly supported by the Department of Energy
under DOE Grant No. DE-FG02-88ER25065. A.A. was supported
by the Swedish Natural Science Research Council (NFR) under
the contract F-FU 06821-304. The work of H.G. was done
in the framework of the project P8916-PHY of the 'Fonds zur
F\"{o}rderung der wissenschaftlichen Forschung in
\"{O}sterreich'.

\end{document}